\documentclass[11pt]{article}
\title{\bf Heat and Fluctuations from\\ Order to Chaos}
\author{\small\textsc{Giovanni Gallavotti}%
\thanks{
Review}\\
\small Dipartimento di Fisica and INFN\\
\small Universit\`a di Roma 
{\it La Sapienza}\\
\small P.~A.~Moro 2, 00185, Roma, Italy\\
\small \texttt{giovanni.gallavotti@roma1.infn.it}
}
\date{\today
}
\newcount\biblio
\font\ottorm=cmr8%

\font\sc=cmcsc10
\def\st{\scriptstyle}%
\font\diecibf=cmbx10%
\font\tenmib=cmmib10 \font\eightmib=cmmib8
\font\sevenmib=cmmib7\font\fivemib=cmmib5 
\font\ottoit=cmti8\font\fiveit=cmti5\font\sixit=cmti6
\font\fivei=cmmi5\font\sixi=cmmi6\font\ottoi=cmmi8%
\font\ottorm=cmr8%
\font\ottosy=cmsy8\font\sixsy=cmsy6\font\fivesy=cmsy5
\font\ottobf=cmbx8\font\sixbf=cmbx6\font\fivebf=cmbx5%
\font\ottocss=cmcsc8%
\def\ottopunti{\def\rm{\fam0\ottorm}\def\it{\fam6\ottoit}\def\em{\fam6\ottoit}%
\def\bf{\fam7\ottobf}%
\textfont1=\ottoi\scriptfont1=\sixi\scriptscriptfont1=\fivei%
\textfont2=\ottosy\scriptfont2=\sixsy\scriptscriptfont2=\fivesy%
\textfont4=\ottocss\scriptfont4=\sc\scriptscriptfont4=\sc%
\textfont5=\eightmib\scriptfont5=\sevenmib\scriptscriptfont5=\fivemib%
\textfont6=\ottoit\scriptfont6=\sixit\scriptscriptfont6=\fiveit%
\textfont7=\ottobf\scriptfont7=\sixbf\scriptscriptfont7=\fivebf%
\setbox\strutbox=\hbox{\vrule height7pt depth2pt width0pt}%
\normalbaselineskip=9pt\rm}
\let\nota=\ottopunti%
\mathchardef\Ba   = "050B  
\mathchardef\Bb   = "050C  
\mathchardef\Bg   = "050D  
\mathchardef\Bd   = "050E  
\mathchardef\Be   = "0522  
\mathchardef\Bee  = "050F  
\mathchardef\Bz   = "0510  
\mathchardef\Bh   = "0511  
\mathchardef\Bthh = "0512  
\mathchardef\Bth  = "0523  
\mathchardef\Bi   = "0513  
\mathchardef\Bk   = "0514  
\mathchardef\Bl   = "0515  
\mathchardef\Bm   = "0516  
\mathchardef\Bn   = "0517  
\mathchardef\Bx   = "0518  
\mathchardef\Bom  = "0530  
\mathchardef\Bp   = "0519  
\mathchardef\Br   = "0525  
\mathchardef\Bro  = "051A  
\mathchardef\Bs   = "051B  
\mathchardef\Bsi  = "0526  
\mathchardef\Bt   = "051C  
\mathchardef\Bu   = "051D  
\mathchardef\Bf   = "0527  
\mathchardef\Bff  = "051E  
\mathchardef\Bch  = "051F  
\mathchardef\Bps  = "0520  
\mathchardef\Bo   = "0521  
\mathchardef\Bome = "0524  
\mathchardef\BG   = "0500  
\mathchardef\BD   = "0501  
\mathchardef\BTh  = "0502  
\mathchardef\BL   = "0503  
\mathchardef\BX   = "0504  
\mathchardef\BP   = "0505  
\mathchardef\BS   = "0506  
\mathchardef\BU   = "0507  
\mathchardef\BF   = "0508  
\mathchardef\BPs  = "0509  
\mathchardef\BO   = "050A  
\mathchardef\BDpr = "0540  
\mathchardef\Bstl = "053F  

\newdimen\xshift \newdimen\xwidth \newdimen\yshift \newdimen\ywidth

\def\ins#1#2#3{\vbox to0pt{\kern-#2pt\hbox{\kern#1pt#3}\vss}\nointerlineskip}

\def\eqfig#1#2#3#4#5{%
\par\xwidth=#1pt\xshift=\hsize\advance\xshift%
by-\xwidth\divide\xshift by 2%
\yshift=#2pt\divide\yshift by 2%
{\hglue\xshift \vbox to #2pt{\vfil%
#3\includegraphics{#4}%
}\hfill\raise\yshift\hbox{#5}}}%

\let\a=\alpha \let\b=\beta  \let\g=\gamma  \let\d=\delta \let\e=\varepsilon
\let\z=\zeta  \let\h=\eta    \let\k=\kappa \let\l=\lambda
\let\m=\mu    \let\n=\nu    \let\x=\xi     \let\p=\pi    \let\r=\rho
\let\s=\sigma \let\t=\tau   \let\f=\varphi 
  \let\ps=\psi   \let\o=\omega
 \let\D=\Delta  \let\L=\Lambda 
    \let\Si=\Sigma     \let\Ps=\Psi
\let\O=\Omega 

\def\*{\vglue3truemm}
\let\0=\noindent
\def\otto{\,{\kern-1.truept\leftarrow\kern-5.truept\to\kern-1.truept}\,}
\def\media#1{{\langle#1\rangle}}
\def\defi{\,{\buildrel def\over=}\,}
\def\EE{{\cal E}}\def\NN{{\cal N}}\def\CC{{\cal C}}
\def\V#1{{\bf#1}}\def\wt#1{{\widetilde#1}}
\def\tende#1{\,\vtop{\ialign{##\crcr\rightarrowfill\crcr
 \noalign{\kern-1pt\nointerlineskip} \hskip3.pt${\scriptstyle
 #1}$\hskip3.pt\crcr}}\,}
\def\T#1{{#1_{\kern-3pt\lower7pt\hbox{$\widetilde{}$}}\kern3pt}}
\def\W#1{#1_{\kern-3pt\lower7.5pt\hbox{$\widetilde{}$}}\kern2pt\,}
\def\lis#1{\overline#1}
\def\wt#1{\widetilde#1}
\def\bra#1{{\langle#1}}\def\ket#1{{|#1\rangle}}
\def\brav#1{{\langle#1|}}
\def\AA{{\cal A}}\def\DD{{\cal D}}\def\PP{{\cal P}}
\def\Im{{\rm Im}\,}
\def\be{\begin{equation}}\def\ee{\end{equation}}
\renewcommand{\theequation}{\arabic{section}.\arabic{equation}}
\def\LL{{\cal L}}

\usepackage{makeidx}         
\usepackage{eqalignno}

\usepackage{fancyhdr}
\def\iniz{\setcounter{equation}{0}{
\rhead{\thepage}\lhead{{{{\diecibf\thesection:}\ \SEC}}}}}

\begin{document}
\maketitle

\begin{abstract}{\it The Heat theorem reveals the second law of
  equilibrium Thermodynamics (i.e.existence of Entropy) as a
  manifestation of a general property of Hamiltonian Mechanics and of
  the Ergodic Hypothesis, valid for $1$ as well as $10^{23}$ degrees
  of freedom systems, {\it i.e.}  for simple as well as very complex
  systems, and reflecting the Hamiltonian nature of the microscopic
  motion. In Nonequilibrium Thermodynamics theorems of comparable
  generality do not seem to be available. Yet it is possible to find
  general, model independent, properties valid even for simple chaotic
  systems ({\it i.e.}  the hyperbolic ones), which acquire special
  interest for large systems: the Chaotic Hypothesis leads to the
  Fluctuation Theorem which provides general properties of certain
  very large fluctuations and reflects the time-reversal symmetry.
  Implications on Fluids and Quantum systems are briefly hinted. The
  physical meaning of the Chaotic Hypothesis, of SRB distributions and
  of the Fluctuation Theorem is discussed in the context of their
  interpretation and relevance in terms of Coarse Grained Partitions
  of phase space. This review is written taking some care that each section
  and appendix is readable either independently of the rest or with
  only few cross references.}\end{abstract} \*
\vfill\eject

\tableofcontents

\kern3mm
{\hbox{\large\bf \kern0mm 19 References\kern93.3mm\pageref{secRef}}}
\vfill\eject

\def\SEC{Heat Theorem}
\section{The Heat Theorem}
\label{sec1}\iniz

An important contribution of Boltzmann to Physics as well as to
research methods in Physics has been the {\it Heat Theorem}.

Summarizing here an intellectual development, spanning about twenty
years of work, the {\it Heat Theorem} for systems of particles of
positions $\V q$ and momenta $\V p$, whose dynamics is modeled by a
Hamiltonian of the form $H=K(\V p)+ W(\V q)$, $K=\frac1{2m}\V p^2$, can
be formulated as follows
\*

\0{\bf Heat Theorem:} {\it In a isolated mechanical system, 
time averages $\media{F}$ of the observables,
{\it i.e.}  of functions $F$ on phase space, are computable as their
integrals with respect to probability distributions $\m_\Ba$ which
depend on the control parameters $\Ba$ determining the states. It is
possible to find four observables, whose averages can be called
$U,V,T,p$, depending on $\Ba$, so that an infinitesimal change $d\Ba$
implies variations $dU,dV$ of $U,V$ so related that

\be \frac{dU+p\, dV}{T}=\,\hbox{``exact''}\,\defi dS\label{1.1}\ee
where $p=\media{-\partial_V W}$ and $V$ is a(ny) parameter on which $W$
depends, and  $U,T$ are the average total energy and the average
total kinetic energy.
\\
When the system is large and $V$ is the volume available to
the particles the quantity $p$ can be shown to have the interpretation
of physical ``pressure'' on the walls of the available volume.}
\*

\0{\it Remarks:} (a) Identification of $T$ with the average kinetic energy
had been for Boltzmann a starting point, assumed a priori, from the
works of Kr\"onig and Clausius of a few years earlier (all apparently
unaware, as everybody else, of the works of Bernoulli, Herapath,
Waterstone, \cite{Br03}).
\\
(b) Connection with observations is made by identifying curves in
parameter space, $t\to\Ba(t)$, with {\it reversible processes}. And in
an infinitesimal process, defined by a line element $d\Ba$, the
quantity $pdV$ is identified with the work the system performs, $dU$
with the energy variation and $dQ=dU+pdV$ as the heat absorbed. Then
relation Eq.(\ref{1.1}) implies that Carnot machines have the highest
efficiency. The latter is one of the forms of the second law, which
leads to the existence of entropy as a function of state in macroscopic
Thermodynamics, \cite{Ze68}.
\\
(c) Eq.(\ref{1.1}), combined with the (independent) assumption that
heat extracted at a fixed temperature cannot be fully transformed into
work, implies that in any process $\frac{dQ}T\le dS$. Hence in
isolated systems changing equilibrium state cannot make entropy
decrease, or in colorful language the {\it entropy of the Universe
cannot decrease}, \cite[p. I-44-12]{Fe63}. Actually by suitably defining
what is meant by irreversible process it is possible to reach the
conclusion that, unless the change of equilibrium state is achieved
via a reversible process, the entropy of an isolated system does
increase strictly, \cite{Ze68}. Conceptually, however, this is an
addition to the second law, \cite[p. I-44-13]{Fe63}.  \*

Examples of control parameters are simply $U,V$, or $T,V$, or $p,V$.
The theorem holds under some hypotheses which evolved from 
\*

\0(a) all motions are periodic (1866) 
\\
(b) aperiodic motions can be considered periodic with infinite
period (!), \cite{Bo66}.
\\
(c) motion visits all phase space of given total energy: in modern
terminology this is the {\it ergodic hypothesis} (1868-1884),
\cite{Bo884}.  \*

The guiding idea was that Eq.(\ref{1.1}) would be true for all systems
described by a Hamiltonian $H=K+W$: {\it no matter whether having few
or many degrees of freedom}, as long as the ergodic hypothesis could
be supposed true.

In other words Eq.(\ref{1.1}) should be considered as a consequence of
the Hamiltonian nature of motions: it is true for all systems
whether with one degree of freedom (as in the 1866 paper by Boltzmann)
or with $10^{19}$ degrees of freedom (as in the 1884 paper by Boltzmann).

It is, in a sense, a property of the particular Hamiltonian structure
of Newton's equations (Hamiltonian given as sum of kinetc plus
potential energy with kinetic energy equal to $\sum_i\frac12{\V
p}_i^2$ and potential energy purely positional). True for all
(ergodic) systems: trivial for $1$ degree of freedom, a surprising
curiosity for few degrees and an important law of Nature for $10^{19}$
degrees of freedom (as in $1\,cm^3$ of ${\rm H_2}$).

The aspect of Boltzmann's approach that will be retained here is that
some universal laws merely reflect basic properties of the equations
of motion which may have deep consequences in large systems: the roots
of the second Law can be found, \cite{Bo66}, in the simple properties
of the pendulum motion.

Realizing the mechanical meaning of the second law induced the
birth of the theory of ensembles, developed by Boltzmann between
1871 (as recognized by Gibbs in the introduction to his treatise) and
1884, hence of Statistical Mechanics.

Another example of the kind are the reciprocal relations of Onsager,
which reflect time reversal symmetry of the Hamiltonian systems
considered above. Reciprocity relations are a first step towards
understanding non equilibrium properties. They impose strong
constraints on transport coefficients, {\it i.e.}  on the $\V
E$-derivatives of various average currents induced by external
forces of intensities $\V E=(E_1,\ldots,E_n)$, which disturb the system
from an equilibrium state into a new {\it stationary
state}. The derivation leads to the quantitative form of reciprocity
which is expressed by the ``Fluctuation-Dissipation Theorems'', {\it
i.e.}  by the Green-Kubo formulae, expressing the transport
coefficient of a current in terms of the mean square fluctuations of
its long time averages.   

In the above Boltzmann's papers (as well as in several other of his
works) Thermodynamics is derived on the assumption that motions are
periodic, hence very regular: see the above mentioned ergodic
hypothesis. Nevertheless heat is commonly regarded as associated with
the chaotic motions of molecules and thermal phenomena are associated
with fluctuations due to chaotic motions at molecular level. A theme
that is pursued in this paper it to investigate how to reconcile
opposites like order and chaos within a unified approach so general to
cover not only equilibrium Statistical Mechanics, but many aspects of
nonequilibrium stationary states. An overview is in the first thiteen
sections, while the appendices enter into technical details, still
keeping at a heuristic level in discusing a matter that is often given
little conideration by Physicists because of its widespread reputation
of being just abstract Mathematics: hopefully this will help to
divulge a theory which is not only simple conceptually nut it seems
promising of further developments. 

The above comment is meant also to explain the meaning of the title of
this paper.
\*

\def\SEC{Time Reversal Symmetry}
\section{Time Reversal Symmetry}
\label{sec2}\iniz
\*

In a way transport coefficients are still equilibrium properties and
nothing is implied by reciprocity when $\V E$ is strictly $\ne\V 0$.

It is certainly interesting to investigate whether time reversal has
important implications in systems which are really out of equilibrium,
{\it i.e.}  subject to non conservative forces which generate currents
(transporting mass, or charge, or heat or several of such
quantities).

There have been many attempts in this direction: it is important to
quote the reference \cite{BK81a} which summarizes a series of works by
a Russian school and completes them. In this paper an extension of the
Fluctuation-Dissipation theorem, as a reflection of time reversal, is
presented, deriving relations which, after having been further
developed, have become known as ``work theorems'' and/or ``transient
fluctuation theorems'' for transformations of systems out of
equilibrium, \cite{Ja97,Ja99,ES94,Cr99,CG99,Ga06b}.

For definiteness it is worth recalling that a dynamical system with
equations $\dot x= f(x)$ in phase space, whose motions will be given 
by maps $t\to S_t x$, is called ``reversible'' if there is a
smooth ({\it i.e.} continuously differentiable) isometry $I$ of phase space,
anticommuting with $S_t$ and involutory, {\it i.e.}

\be I S_t=S_{-t}I,\qquad I^2=1\label{2.1}\ee
Usually, if $x=(\V p,\V q)$, time reversal is simply $I(\V p,\V q)=(-\V
p,\V q)$.

The main difficulty in studying nonequilibrium statistical Mechanics
is that, after realizing that one should first understand the
properties of stationary states, considered as natural extensions of the
equilibrium states, it becomes clear that the microscopic description
{\it cannot be Hamiltonian}.

This is because a current arising from the action of a nonconservative
force continuously generates ``heat'' in the system. Heat has to be
taken out to allow reaching a steady state. This is empirically done
by putting the system in contact with one or more thermostats. In
models, thermostats are just forces which act performing work
balancing, at least in average, that produced by the external
forces, {\it i.e.}  they ``model heat extraction''.

It is not obvious how to model a thermostat; and any thermostat model
is bound to be considered ``unphysical'' in some respects. This is not
surprising, but it is expected that most models introduced to describe
a given physical phenomenon should be ``equivalent''. 

Sometimes it is claimed that the only physically meaningful
thermostats for nonequilibrium systems (in stationary states) are made
by infinite ($3$-dimensional) systems which, asymptotically at
infinity, are in statistical equilibrium. In the latter cases it is
not even necessary to introduce {\it ad hoc} forces to remove the
heat: {\it motion remains Hamiltonian} and heat flows towards
infinity.

Although the latter is certainly a good and interesting model, as
underlined already in \cite{FV63}, it should be stressed that it is
mathematically intractable unless the infinite systems are
``free''. {\it i.e.}  without internal interaction other than linear,
\cite{FV63,HL73b,EPR99,APJP06,HI05}. 

And one can hardly consider such assumption more physical than the one
of finite thermostats.  Furthermore it is not really clear whether a linear
external dynamics can be faithful to Physics, as shown by the simple
one dimensional XY-models, see \cite{ABGM72} where a linear thermostat
dynamics with a single temperature leads a system to a stationary
state, as expected, but the state is not a Gibbs state (at any
temperature). The method followed in \cite{ABGM72}, based on
\cite{BD69}, can be used to illustrate some problems which can arise
when thermostats are classical free systems, see Appendix A4.

\def\SEC{Point of view}
\section{Point of view}
\label{sec3}\iniz

The restriction to finite thermostats, followed here, is not chosen
because  infinite thermostats should be considered
unphysical, but rather because it is a fact that the recent progress in
nonequilibrium theory can be traced to
\*

\0(a) the realization of the interest of restricting attention 
to {\it stationary states}, or {\it
steady states}, reached under forcing (rather than discussing approach
to equilibrium, or to stationarity).

\0(b) the simulations on steady states performed in the 80's after the
essential role played by {\it finite thermostats} was fully realized.
\*

Therefore investigating finite thermostat models is still particularly
important.  This makes in my view interesting to confine attention on
them and to review their conceptual role in the developments that took
place in the last thirty years or so.

Finite thermostats can be modeled in several ways: but in constructing
models it is desirable that the models keep as many features as
possible of the dynamics of the infinite thermostats. As realized in
\cite[p.452]{BK81a} it is certainly important to maintain the {\it
time reversibility}. Time reversibility expressed by Eq. (\ref{2.1}), {\it
i.e.}  existence of a smooth conjugation between past and future,
is a fundamental symmetry of nature which (replaced by TCP) even
``survives'' the so called time reversal violation; hence it is desirable
that it is saved in models. An example will be discussed later.
\*

\0{\it Comment:} (1) The second law of equilibrium Thermodynamics,
stating existence of the state function entropy, can be derived
without reference to the microscopic dynamics by assuming that heat
absorbed at a single temperature cannot be cyclically converted into
work, \cite{Ze68}. In statistical Mechanics equilibrium, states are
identified with probability distributions on phase space: they depend
on control parameters (usually two, for instance energy and volume)
and processes are identified with sequences of equilibrium states,
{\it i.e.} as curves in the parameters space interpreted as {\it
reversible processes}. The problem of how the situation, in which
averages are represented by a probability distribution, develops
starting from an initial configuration {\it is not part of the
equilibrium theory}. In this context the second law arises as a
theorem in Mechanics (subject to asssumptions) and, again, just says
that entropy exists (the heat theorem).
\\
(2) As noted in Sec.\ref{sec1}, if the scope of the theory is enlarged
admitting processes that cannot be represented as sequences of
equilibria, called ``irreversible processes'', then the postulate of
impossibility to convert heat into work extracting it from a single
thermostat implies, again without involving microscopic dynamics, the
inequality often stated as ``the entropy of the Universe'' cannot
decrease in passing from an equilibrium state to another. And, after
properly defining what is meant by irreversible process \cite{Ze68},
actually strictly increases if in the transformation an irreversible
process is involved; however perhaps it is best to acknowledge
explicitly that such a strict increase is a further assumption,
\cite[p. I-44-13]{Fe63} leaving aside a lengthy, \cite{Ze68}, and
possibly not exhaustive analysis of how in detail an irreversible
transformation looks like. Also this second statement, under suitable
assumptions, can become a theorem in Mechanics, \cite{Le93,Ga01}, but
here this will not be discussed.
\\
(3) Therefore studying macroscopic properties for systems out of
equilibrium can be divided into an ``easier'' problem, which is the
proper generalization of equilibrium statistical Mechanics: namely
studying stationary states identified with corresponding probability
distributions yielding, by integration, the average values of the few
observables of relevance. And the problem of approach to a stationary
state which is of course more difficult. The recent progress in
nonequilibrium has been spurred by restricting research to the easier
problem.

\def\SEC{Chaotic Hypothesis}
\section{The Chaotic Hypothesis (CH)}
\label{sec4}\iniz

Following Boltzmann and Onsager we can ask whether there are general
relations holding among time averages of selected observables and for
all systems that can be modeled by time reversible mechanical
equations $\dot x=f(x)$.

The difficulty is that in presence of dissipation it is by no means
clear which is the probability distribution $\m_\Ba$ which provides
the average values of observables, at given control parameters $\Ba$.

In finite thermostat models dissipation is manifested by the
nonvanishing of the divergence, $\s(x)\defi-\sum \partial_{x_i}
f_i(x)$, of the equations of motion and of its time average $\s_+$.

If $\s_+>0$%
\footnote{\nota As intuition suggests $\s_+$ cannot be $<0$,
\cite{Ru97}, when motion takes place in a bounded region of phase
space, as it is supposed here.}, it is not possible that the
distributions $\m_\Ba$ be of the form $\r_\Ba(x)dx$, ``absolutely
continuous with respect to the phase space volume'': since volume
contracts, the probability distributions that, by integration, provide
the averages of the observables must be concentrated on sets,
``attractors'', of $0$ volume in phase space.

This means that there is no obvious substitute of the ergodic
hypothesis: which, however, was essential in equilibrium statistical
Mechanics to indicate that the ``statistics'' $\m_\Ba$, {\it i.e.}  the
distribution $\m_\Ba$ such that

\be \lim_{T\to\infty}\frac1T\int_0^T F(S_tx)dt=\int \m_\Ba(dy)
F(y)\label{4.1}\ee
for all $x$ except a set of zero volume, exists and is given by the
Liouville volume (appropriately normalized to $1$) on the surfaces of
given energy $U$ (which is therefore one of the parameters $\Ba$ on
which the averages depend).\footnote{\nota By Liouville volume we mean
the measure $\d(K(\V p)+ W(\V q)-U)d\V p d\V q$, on the manifold of
constant energy or, in dissipative cases discussed later, the measure
$d\V p d\V q$.\label{2}\label{footnote2}}

It is well known that identifying $\m_\Ba$ with the Liouville volume
does not allow us to derive the values of the averages (aside from a
few very simple cases, like the free gas): but it allows us to write
the averages as explicit integrals, \cite{Ga00}, which are well suited to
deduce relations holding between certain averages, like the second law
Eq.(\ref{1.1}) or Onsager reciprocity and the more general Fluctuation
Dissipation Theorems.

The problem of finding a useful representation of the statistics of
the stationary states in systems which are not in equilibrium arose in
the more restricted context of fluid Mechanics earlier than in
statistical Mechanics. And through a critique of earlier attempts,
\cite{Ru78b}, in 1973 Ruelle proposed that one should take advantage of
the empirical fact that motions of turbulent systems are ``chaotic''
and suppose that their mathematical model should be a ``hyperbolic
system'', in the same spirit in which the ergodic hypothesis should be
regarded: namely {\it while one would be very happy to prove
ergodicity because it would justify the use of Gibbs' microcanonical
ensemble, real systems perhaps are not ergodic but behave nevertheless
in much the same way and are well described by Gibbs' ensemble...},
\cite{Ru73}.

The idea has been extended in \cite{GC95,Ga00} to nonequilibrium statistical
Mechanics in the form

\* \0{\bf Chaotic hypothesis (CH):} {\it Motions on the attracting set
of a chaotic system can be regarded as motions of a smooth transitive
hyperbolic system.\footnote{\nota Transitive means ``having a dense
orbit''. Note that here this is a property of the attracting set,
which is often not at all dense in the full phase space.  Such systems
are also called ``Anosov systems''.}}  \*

The hypothesis was formulated to explain the result of the experiment
in \cite{ECM93}. In \cite{GC95} it was remarked that the CH could be
adequate for the purpose.

\*
\def\SEC{``Free'' implications of CH}
\section{``Free'' implications of the Chaotic Hypothesis}
\label{sec5}\iniz

Smooth transitive hyperbolic systems share, independently of the number
of degrees of freedom, remarkable properties, \cite{GBG04}.  \*

\0(1) their motions can be considered paradigmatic chaotic evolutions,
    whose theory is, nevertheless, very well understood to the point
    that they can play for chaotic motions a role alike to the one
    played by harmonic oscillators for ordered motions, \cite{Ga95a}.

\0(2) there is a {\it unique} distribution $\m$ on phase space such that 

\be \lim_{\t\to\infty}\frac1\t\int_0^\t F(S_tx)dt=\int \m(dy) F(y)
\label{5.1}\ee
for all smooth $F$ and for all but a zero volume set of initial data
$x$, \cite{Si72,BR75,Ga00,GBG04}, see Appendix A1. The distribution
$\m$ is called the {\it SRB probability distribution}, see Appendix A2.

\0(3) averages satisfy a {\it large deviations rule}:
{\it i.e.} if the point $x$ in $f=\frac1\t\int_0^\t F(S_tx)\,dt$ 
is sampled with distribution $\m$, then

\be \lim_{\t\to\infty} 
\frac1\t\log Prob_\m(f\in\D)= \max_{f\in\D} \z_F(f)\label{5.2}\ee
is an asymptotic value that controls the probability that the finite
time average of $F$ falls in an interval $\D=[u,v], \,u<v$, subset of
the interval $(a_F,b_F)$ of definition of $\z_F$. In the interval of
definition $\z_F(f)$ is convex and analytic in $f$, \cite{Si72,Si77}.
Outside $[a_F,b_F]$ the function $\z_F(f)$ can be defined to have
value $-\infty$ (which means that values of $f$ in intervals outside
$[a_F,b_F]$ can possibly be observed only with a probability tending
to $0$ faster than exponentially), \cite{Si72,Si77}.

\0(4) A more precise form of Eq.(\ref{5.2}) yields also the rate at
which the limit is reached: $Prob_\m(f\in\D)= e^{\t\,\max_{f\in\D}
\z_F(f)+ O(1)}$ with $O(1)$ bounded uniformly in $\t$, at fixed
distance of $\D$ from the extremes $a_F,b_F$. This is ofteen written
in a not very precise but mnemocnically convenient form, as long as
its real meaning is kept in mind, as

\be P_\m(f)=e^{\t\,\z_F(f)+O(1)}\label{5.3}\ee

\0(5) The fluctuations described by (\ref{5.2}) are very large
fluctuations as they have size of order $\t$ rather than
$O(\sqrt{\t})$: in fact if the maximum of $\z_F(f)$ is at a point
$f_0\in(a_F,b_F)$ and is a nondegenerate quadratic maximum, then
Eq. (\ref{5.2}) implies that $\sqrt{\t}(f-f_0)$ has an asymptotically
Gaussian distribution. This means that the motion can be regarded to be so
chaotic that the values of $F(S_tx)$ are independent enough so that
the finite time average deviations from the mean value $f_0$ are
Gaussian on the scale of $\sqrt\t$.

\0(6) A natural extension to (\ref{5.2}) in which several observables
$F_1,\ldots,F_n$ are simultaneously considered is obtained by defining
$f_i=\frac1\t\int_0^\t F_i(S_tx)dt$. Then there exists a convex closed
set $C\subset {\cal R} ^n$ and function $\z_{\V F}(\V f)$ analytic in $\V
f=(f_1,\ldots,f_n)$ in the interior of $C$ and, given an open set
$\BD\subset C$,

\be \lim_{\t\to\infty} \frac1\t\log Prob_\m(\V f\in\BD)= \max_{\V
f\in\BD}\,\,\z_{\V F}(\V f)\label{5.4}\ee
and $\z_{\V F}(\V f)$ could be defined as $-\infty$ outside $C$, with
the meaning mentioned in remark (2). If the function $\z_{\V F}(\V f)$
attains its maximum in a point $\V f_0$ in the interior of $C$ and the
maximum is quadratic and nondegenerate, then the {\it joint
fluctuations} of $\Bf=\sqrt\t(\V f-\V f_0)$ are asymptotically
Gaussian, which means that have a probability density
$\frac1{\sqrt{\p^n\det \DD}} e^{-\frac12 (\Bf\cdot \DD^{-1}\Bf)}$ with
$\DD$ a positive definite $n\times n$ matrix.

\0(7) The probability distribution $\m$ depends on the control
    parameters $\Ba$ of the initial data and therefore as $\Ba$ varies
    one obtains a collection of probability distributions: this leads
    to a natural {\it extension of the ensembles of equilibrium
    statistical Mechanics}, \cite{Ga00}.

\0(8) The most remarkable property, root of all the above, is that the
    SRB probability distribution $\m$, can be given a concrete formal
    representation, in spite of being a distribution concentrated on a
    set of zero volume, \cite{Si72,Si77}, see Appendix A1,A2. This raises
    hopes to use it to derive general relations between averages of
    observables. As in equilibrium, the averages with respect to $\m$
    are destined to remain not computable except, possibly, under
    approximations (aside very few exactly soluble cases): their
    formal expressions could nevertheless be used to establish general
    mutual relations and properties.

\0(9) Given the importance of the existence and representability of
    the SRB distribution, Appendix A1,A2 will be entirely devoted to
    the formulation (A1) and to the physical interpretation of the
    derivation of its expression: this could be useful for readers who
    want to understand the technical aspects of what follows, because
    some may find not satisfactory skipping the technical details even
    at a heuristic level. The aim of the non technical discussion that
    follows, preceding the appendices, is to make it worth to invest
    some time on the technical details.

\0(10) Applied to a system in equilibrium the CH implies the ergodic
    hypothesis so that it is a genuine extension of the latter and any
    results that follow from it will be necessarily compatible with
    those of equilibrium statistical Mechanics, \cite{Ga00}.

\0(11) For very simple systems the distribution $\m$ can be constructed
    explicitly and time averages of some observables computed. The
    systems are the discrete time evolutions corresponding to linear
    hyperbolic maps of tori, \cite{GBG04}, or the continuous time geodesic
   motion on a surface of constant negative curvature. The latter
    systems are rigorously hyperbolic and the SRB distribution can be
    effectively computed for them {\it as well as for their small
    perturbations}.

\0(12) A frequent remark about the chaotic hypothesis is that it does
    not seem to keep the right viewpoint on nonequilibrium
    Thermodynamics. It should be stressed that the hypothesis is
    analogous to the ergodic hypothesis, which ({\it as well known})
    cannot be taken as the foundation of equilibrium statistical
    Mechanics, even though it leads to the correct Maxwell Boltzmann
    statistics, because the latter ``holds for other reasons''.
    Namely it holds because in most of phase space (measuring sizes by
    the Liouville measure) the few interesting macroscopic observables
    have the same value, \cite{Th74}, see also \cite{Le93}.  \*

\def\SEC{Paradigms}
\section{Paradigms of Statistical Mechanics and CH}
\label{sec6}\iniz

    In relation to the last comment is useful to go back to the Heat
    Theorem of Sec.1 and to a closer examination of the basic paper of
    Boltzmann \cite{Bo884}, in which the theory of equilibrium
    ensembles is developed and may offer arguments for further
    meditation.  The paper starts by illustrating an important, and
    today almost forgotten, remark by Helmoltz showing that very
    simple systems (``monocyclic systems'') can be used to construct
    mechanical models of Thermodynamics: and the example chosen by
    Boltzmann is {\it really extreme by all standards}.

    He shows that the motion of a {\it Saturn ring} of mass $m$ on
    Keplerian orbits of major semiaxis $a$ in a gravitational field of
    strength $g$ can be used to build a model of Thermodynamics.  In
    the sense that one can call \
\*
\0``volume'' $V$ the gravitational
    constant $g$,
\\
``temperature'' $T$ the average kinetic energy,
\\
    ``energy'' $U$ the energy and
\\
 ``pressure'' $p$ the average potential
    energy $m k a^{-1}$,
\*
\0then one infers that by varying, {\it at fixed
    eccentricity}, the parameters $U,V$ the relation
    $(dU+pdV)/T=\,\hbox{\it exact}$ holds.  Clearly this {\it could}
    be regarded as a curiosity, see \cite[Appendix 1.A1, Appendix 9.A3]{Ga00}.

    However Boltzmann (following Helmoltz?\footnote{\nota The relation
    between the two on this subject should be more
    studied. Boltzmann's paper of 1884, \cite{Bo884}, is a natural
    follow up and completion of his earlier work \cite{Bo871b} which
    followed \cite{Bo868,Bo66}. It seems that the four extremely long
    papers by Helmoltz, also dated 1884, \cite{He884a,He884b}, might
    have at most just stimulated Boltzmann to revisit his earlier
    works and led him achieve the completion of the mechanical
    explanation of the second law. Certainly Boltzmann attributes a
    strong credit to Helmoltz, and one wonders if this might be partly
    due to the failed project that Boltzmann had to move to Berlin
    under the auspices of Helmoltz.}) took it seriously and proceeded
    to infer that under the ergodic hypothesis {\it any system} small
    or large provides us with a model of Thermodynamics (being
    ``monocyclic'' in the sense of Helmoltz): for instance he showed
    that the canonical ensemble verifies exactly the second law of
    equilibrium Thermodynamics (in the form $(dU+p\,dV)/T=\, \hbox{\it
    exact}$) {\it without any need to take thermodynamic limits},
    \cite{Bo884}, \cite{Ga00}.  The same could be said of the
    microcanonical ensemble (here, however, he had to change
    ``slightly'' the definition of heat to make things work without
    finite size corrections).

    He realized that the Ergodic Hypothesis could not possibly account
    for the correctness of the canonical (or microcanonical)
    ensembles; this is clear at least from his (later) paper in
    response to Zermelo's criticism, \cite{Bo96}. Nor it could account
    for the observed time scales of approach to equilibrium.
    Nevertheless he called the theorem he had proved the {\it heat
    theorem} and never seemed to doubt that it provided evidence for
    the correctness of the use of the equilibrium ensembles for
    equilibrium statistical Mechanics.

    Hence there are two points to consider: first certain
    relations among mechanical quantities {\it hold no matter how
    large} is the size of the system and, secondly, they can be seen
    and tested not only in small systems, by direct measurements, but
    even in large systems, because in large systems such mechanical
    quantities acquire a macroscopic thermodynamic meaning and their
    relations are ``typical'' {\it i.e.}  they hold in most of phase space.

    The first point has a close analogy in that the consequences of
    the Chaotic Hypothesis stem from the properties of small dimension
    hyperbolic systems (the best understood) which play here the role
    of Helmoltz' monocyclic systems of which Boltzmann's Saturn ring
    (\cite{Bo884}) is a special case.  They are remarkable
    consequences because they provide us with {\it parameter free
    relations} (namely the Fluctuation Theorem, to be discussed below,
    and its consequences): but clearly it cannot be hoped that a
    theory of nonequilibrium statistical Mechanics be founded solely
    upon them, by the same reasons why the validity of the second law
    for monocyclic systems had in principle no reason to imply the
    theory of ensembles.

    Thus what is missing are arguments similar to those used by
    Boltzmann to justify the use of ensembles, {\it independently} of
    the ergodic hypothesis: an hypothesis which in the end may appear
    (and still does appear to many) as having only suggested them ``by
    accident''.  The missing arguments should justify the CH on the
    basis of an extreme likelihood of its predictions in systems that
    are very large and that may be not hyperbolic in the mathematical
    sense.  I see no reason, now, why this should prove impossible
    {\it a priori} or in the future.  See Sect.\ref{sec12} for some of
    the difficulties that can be met in experiments testing the CH
    through its consequence discussed in Sec.\ref{sec7}.

\* In the meantime it seems interesting to take the same philosophical
    viewpoint adopted by Boltzmann: not to consider a chance that {\it
    all} chaotic systems share some selected, and remarkable,
    properties and try to see if such properties help us achieving a
    better understanding of nonequilibrium. After all it seems that
    Boltzmann himself took a rather long time to realize the interplay
    of the just mentioned two basic points behind the equilibrium
    ensembles and to propose a solution harmonizing them. ``All it
    remains to do'' is to explore if the hypothesis has implications
    more interesting or deeper than the few known and presented in the
    following.

\def\SEC{Fluctuation Theorem}
\section{The Fluctuation Theorem (FT)}
\label{sec7}\iniz

The idea of looking into time reversibility to explain the
experimental results of \cite{ECM93} is clearly expressed in the same
paper.  The CH allows us to use effectively time reversal symmetry
to obtain what has been called in \cite{GC95,Ga95b,GC95b} the ``{\it
Fluctuation Theorem}''. In fact a simple property holds for all
transitive hyperbolic systems which admit a time reversal symmetry.

The property deals with the key observable $\s(x)$, which is the above
introduced divergence of the equations of motion, or ``phase space
contraction rate''. Assuming the average phase space contraction to be
positive, $\s_+>0$, let $p=\frac1\t\int_0^\t \frac{\s(S_t x)}{\s_+}dt$
be the ``dimensionless phase space contraction''; let $\z(p)$ be the large
deviation rate function introduced in \S\ref{sec5}, see
Eq.(\ref{5.2}), for $F(x)=\frac{\s(x)}{\s_+}$. By time reversal
symmetry the interval of analyticity of $\z(p)$ is centered at the
origin and will be denoted $(-p^*,p^*)$; furthermore $p^*\ge1$,
because the average of $p$ is $1$. Then, \cite{GC95},

\*
\0{\bf Fluctuation Theorem (FT):} 
{\it The probabilities of the large deviations of $p$ satisfy, for all
transitive time reversible hyperbolic systems, 

\be \z(-p)=\z(p)-p\s_+\label{7.1}\ee
for all $|p|<p^*$: this will be called a ``fluctuation relation'',
(FR).  } \*

\0{\it Remarks:} 
\\
(1) In terms of the notation in Eq.(\ref{5.3}) the FT is 

\be\frac{P_\t(p)}{P_\t(-p)}=e^{\,p\,\s_+\,\t +O(1)}\label{7.2}\ee
which is the form in which it is often written.
\\
(2) The theorem has been developed, in \cite{GC95}, to understand the
results of a simulation, \cite{ECM93}, whose Authors had correctly
pointed out that the SRB distribution together with the time
reversibility could possibly explain the observations.
\\
(3) Unfortunately the same name, introduced in \cite{GC95,Ga95b,GC95b}
where FT has been proved, has been {\it subsequently} picked up and
attributed to other statements, superficially related to the above
FT. Enormous confusion ensued (and sometimes even errors), see
\cite{CG99,GC04,GZG05}. A more appropriate name for such other, and
different, statements has been suggested to be ``transient fluctuation
theorems''.  The above FT should be distinguished also from the
results in \cite{BK81a} which were the first {\it transient}
fluctuations results, later extended and successfully applied,
see \cite{Ja97,Ja99}. It is claimed that the difference between the
above FT and the transient statements is just an exchange of limits:
the point is that it is a nontrivial one, see counterexamples in
\cite{CG99}, and assumptions are needed, which have a physical
meaning; the CH is the simplest.
\\
(4) The FT theorem has been proved first for discrete time evolutions,
{\it i.e.}  for maps: in this case the averages over time are
expressed by sums rather than by integrals. Hyperbolic maps are
simpler to study than the corresponding continuous time systems, which
we consider here, because smooth hyperbolic maps do not have a trivial
Lyapunov exponent (the vanishing one associated with the phase space
flow direction); but the techniques to extend the analysis to
continuous time systems are the same as those developed in \cite{Ge98}
for proving the FT for hyperbolic flows and in this review I shall not
distinguish between the two kinds of evolutions since the properties
considered here do not really differ in the two cases.
\\
(5) The condition $\s_+>0$, {\it i.e.}  dissipativity, is {\it
  essential} even to define $p$ itself. When the forcing intensity $E$
  vanishes also $\s_+\to0$ and the FR loses meaning because $p$
  does. Neverheless by appropriately dividing both sides of
  Eq. (\ref{7.1}) by $\s_+$, and then taking the limit, a nontrivial
  limit can be found and it can be shown, at least heuristically, to
  give the Green-Kubo relation for the ``current''
  $J\defi\media{\frac{\partial\s}{\partial E}}_{\m}=\media{j}_\m$,
  \cite{Ga96a,Ga00}, generated by the forcing, namely

\be \frac{d J}{d E}\Big|_{E=0}=\frac12\int_{-\infty}^\infty \media{j(S_\t
x)\,j(x)}_{E=0}dt\label{7.3}\ee
which is a general Fluctuation-Dissipa\-tion theorem.
\\
(6) The necessity of a bound $p^*$ in FT has attracted undue
    attention: it is {\it obvious} that it is there since $\s(x)$ is
    bounded, if CH holds. It also true that the role of $p^*$ is
    discussed in the paper \cite{Ga95b}, which is a formal and {\it
    contemporary} version of the earlier \cite{GC95} and of part of
    the later \cite{GC95b} written for a different audience in mind.
\\
It is therefore surprising that this is sometimes ignored in the
    literature and the original papers are faulted for not mentioning
    this (obvious) point, which in any event is fully discussed in
    \cite{Ga95b}. A proof which also discusses $p^*$ is in
    \cite{Ru99}. It is also obvious that for $p\ge p^*$ the function
    $\z(p)$ can be naturally set to be $-\infty$, as commented in
    remark (6) to the CH in Sec.\ref{sec4}, and for this reason
    Eq. (\ref{7.1}) is often written without any restiction on $p$.
    This is another point whose misunderstanding has led to
    errors. For readers familiar with statistical Mechanics there is
    nothing misterious about $p^*$.  It is analogous the ``close
    packing density'' in systems with hard cores: it is clear that
    there is a well defined maximum density but its value is not
    always explicitly computable; and for hiher density many
    thermodynamic functions may be considered defined but as having an
    infinite value.  \*

\0{\bf Corollary:} {\cite{Ga97,Ga00},\it Under the same assumptions of FT,
if $F_1=\frac{\s(x)}{\s_+}$, $F_2,\ldots, F_n$ are $n$ observables of
parity $\e_i=\pm$ under time reversal, $F_i(Ix)=\e_i F_i(x)$, the
large deviations rate $\z_{\V F}(\V f)$, defined in Eq. (\ref{5.4}),
satisfies

\be \z_{\V F}(\V f^*)=\z_{\V F}(\V f)-\s_+ f_1\label{7.4}\ee
where $\V f^*=(- f_1,\e_2 f_2,\ldots,\e_n f_n)$, in its domain of definition
$C\subset {\cal R}^n$.}

\*
\0{\it Remark:} Note that the {\it r.h.s.} of Eq.(\ref{7.4}) {\it does not
depend on $f_2,\ldots,f_n$}. The independence has been exploited in
\cite{Ga96a} to show that when the forcing on the system is due to several
forces of respective intensities $E_1,\ldots,E_s$ then by taking $F_1=
\frac{\s(x)}{\s_+},\, F_2=\partial_{E_k}\s(x)$, the Eq.(\ref{7.4}) implies,
setting $j_k(x)=\partial_{E_k}\s(x)$ and $J_k=\media{j_k}_\m$, the Green
Kubo relations (hence Onsager reciprocity)

\be L_{hk}=\partial_{E_h} J_k\big|_{\V E=\V0}=\frac12\int_{-\infty}^\infty
\media{j_h(S_\t
x)\,j_k(x)}_{E=0}dt=L_{kh}.\label{7.5}\ee
Therefore FT can be regarded as an {\it extension} to a nonlinear
regime of Onsager reciprocity and of the Fluctuation-Dissipation
theorems. Such a relation was pointed out in the context of volume
preserving dynamics (hence in absence of dissipation), see comments in
\cite[p.452]{BK81a} in particular. But it is not clear how to obtain
from \cite{BK81a} the dissipative case results in
Eq.(\ref{7.1}),(\ref{7.4}),(\ref{7.5}) without the CH.  \*

\def\SEC{Patterns and Onsager-Machlup Fluctuations}
\section{Fluctuation Patterns, Onsager-Machlup Theory}
\label{sec8}\iniz

The last comment makes it natural to inquire whether there are more
direct and physical interpretations of the FT (hence of the
meaning of CH) when the external forcing is really different from the
value $0$ (the value always assumed in Onsager's theory).

The proof of the FT allows, as well, to deduce, \cite{Ga99}, an
apparently more general statement (closely related to a relation
recently found in the theory of the Kraichnan model of $2$-dimensional
turbulence and called ``multiplicative'' fluctuation theorem,
\cite{CDG07}) which can be regarded as an extension to nonequilibrium
of the Onsager-Machlup theory of fluctuation patterns.

Consider observables $\V F=(F_1\defi {\s}/{\s_+},\ldots,F_n)$ which
have a well defined time reversal parity: $F_i(Ix)=\e_{F_i} F_i(x)$,
with $\e_{F_i}=\pm1$.  Let $F_{i+}$ be their time average ({\it i.e.}
their SRB average) and let $t\to \Bf(t)=(\f_1(t),\ldots,$ $\f_n(t))$
be a smooth bounded function.  Look at the probability, relative to
the SRB distribution ({\it i.e.}  in the ``natural stationary state'')
that $F_i(S_t x)$ is $\f_i(t)$ for $t\in [-\frac\t2,\frac\t2]$: we say
that $\V F$ ``follows the fluctuation pattern'' $\Bf$ in the time
interval $t\in [-\frac\t2,\frac\t2]$.

No assumption on the fluctuation size, nor on the size of the forces
keeping the system out of equilibrium, will be made.  Besides the
CH we assume, however, that the evolution is time
reversible {\it also} out of equilibrium and that the phase space
contraction rate $\s_+$ is not zero (the results hold no
matter how small $\s_+$ is and, appropriately interpreted, they make
sense even if $\s_+=0$, but in that case they become trivial).

We denote $\z(p,\Bf)$ the {\it large deviation function} for observing
in the time interval $[-\frac\t2,\frac\t2]$ an average phase space
contraction $\s_\t\defi\frac1\t\int_{-\t/2}^{\t/2}\s(S_tx)dt= p\s_+$
and at the same time a fluctuation pattern $\V F(S_tx)=\Bf(t)$.  This
means that the probability that the {\it dimensionless phase space
contraction rate} $p$ is in a closed set $\D$ and $F$ is in a closed
neighborhood of an assigned $\Bps$,\footnote{\nota By ``closed
neighborhood'' $U_{\Bps,\e}$, $\e>0$, around $\Bps$, we mean that
$|F_i(S_tx)-\ps_i(t)|\le\e$ for $t\in[-\frac\t2,\frac\t2]$.}  denoted
$U_{\Bps,\,\e}$, is given by:

\be \exp\Big(\sup_{p\in\D,\Bf\in U_{\Bps,\e}}
{\t\,\z(p,\Bf)}\Big)\label{8.1}\ee
to leading order as $\t\to\infty$ ({\it i.e.}  the logarithm of the mentioned
probability divided by $\tau$ converges as $\t\to\infty$ to
$\sup_{p\in\D,\Bf\in U_{\Bps,\e}} \z(p,\Bf)$). Needless to say $p$ and
$\Bf$ have to be ``possible'' otherwise $\z$ has to be set $-\infty$, as
in the FT case in Sec.\ref{sec6}, comment (6).

Given a reversible, dissipative, transitive Anosov flow the
fluctuation pattern $t\to\Bf(t)$ and the time reversed pattern
$t\to\e_F\Bf(-t)$ are then related by the following: 
\*

\0{\bf Conditional reversibility relation:} {\sl If $\V
F=(F_1,\ldots,F_n)$ are $n$ observables with defined time reversal
parity $\e_{F_i}=\pm1$ and if $\t$ is large the fluctuation pattern
$\Bf(t)$ and its time reversal $I\f_i(t)\defi \e_{F_i}\f_i(-t)$ will
be followed with equal likelihood if the first is conditioned to a
contraction rate $p$ and the second to the opposite $-p$.  This
holds because:

\be \frac{\z(p,\Bf)-\z(-p,I\Bf) }{p\s_+}=1 \qquad {\rm for\ } |p|\le
p^*\label{8.2}\ee
with $\z$ introduced in Eq.(\ref{8.1}) and a suitable $p^*\ge1$.}
\*

It will appear, in Sec.\ref{sec9}, that the phase space contraction
rate should be identified with a macroscopic quantity, 
the {\it entropy creation rate}. Then
the last theorem can be interpreted as saying, in other words, that
while it is very difficult, in the considered systems, to see an
``anomalous'' average entropy creation rate during a time $\t$ ({\it
e.g.} $p=-1$), it is also true that ``that is the hardest thing to
see''.  Once we see it, {\it all the observables will behave
strangely} and the relative probabilities of time reversed patterns
will become as likely as those of the corresponding direct patterns
under ``normal'' average entropy creation regime.

``A waterfall will go up, as likely as we expect to see it going down,
in a world in which for some reason the entropy creation rate has
changed sign during a long enough time.'' We can also say that the
motion on an attractor is reversible, even in presence of dissipation,
once the dissipation is fixed.

The result in Eq.(\ref{8.2}) is a ``{\it relation}'' rather than a
theorem because, even in the hyperbolic cases, the precise
restrictions on the ``allowed'' test functions $\f_i(t)$ have not been
discussed in \cite{Ga99} from a strict mathematical viewpoint and it
would be interesting to formulate them explicitly and investigate
their generality.\footnote{\nota A sufficient condition should be that
$\f_i(t)$ are bounded and smooth.}\*

The result can be informally stated in a only apparently
stronger form, for $|p|<p^*$, and with the warnings in remark (4)
preceding the analogous Eq.(\ref{5.3}), as 

\be \frac{P_\t(\hbox{\rm
for all}\ j, {\rm and}\, t\in[-\frac12\t,\frac12\t] \,:
F_j(S_tx)\sim\f_j(t))} {P_\t(\hbox{\rm for all}\ \,j, {\rm and}\,
t\in[-\frac12\t,\frac12\t]\,: F_j(S_tx)\sim -\f_j(-t))}
=e^{\,p\,\s_+\, \t+O(1)},\label{8.3}\ee
where $P_\t$ is the SRB probability, {\it provided the phase space
contraction $\s(x)$ is a function of the observables $\V F$}. This is
certainly the case if $\s$ is one of the $F_i$, for instance if
$\s=F_1$. Here $F_j(S_tx)\sim \f_j(t)$ means
$|F_j(S_tx)-\f_j(t)|$ small for $t\in[-\frac\t2,\frac\t2]$.
\*

\0{\it Remarks:} 
\\
 (1) A relation of this type has been remarked
recently in the context of the theory of Lagrangian trajectories in
the Kraichnan flow, \cite{CDG07}.  
\\
 (2) One should note that in applications results like Eq.(\ref{8.3})
will be used under the CH and therefore other errors may arise because
of its approximate validity (the hypothesis in fact essentially states
that ``things go as if'' the system was hyperbolic): they may depend
on the number $N$ of degrees of freedom and we do not control them
except for the fact that, if present, their relative value should tend
to $0$ as $N\to\infty$: there may be (and there are) cases in which
the chaotic hypotesis is not reasonable for small $N$ ({\it e.g.}
systems like the Fermi-Pasta-Ulam chains) but it might be correct for
large $N$.  We also mention that, on the other hand, for some systems
with small $N$ the CH may be already regarded as valid ({\it e.g.} for
the models in \cite{CELS93}, \cite{ECM93,BGG97}).
\\
(3) The proofs of FT and the corollaries are not difficult. Once their
meaning in terms of coarse graining is understood, the a priori rather
misterious SRB distribution $\m$ is represented, surprisingly, as a
Gibbs distribution for a $1$--dimensional spin system, which is
elementary and well understood. In Appendix A1,A2 some details are given
about the nature of coarse graining and in Appendix A3 the steps of
the proof of FT are illustrated.  
\* 

In conclusion the FT is a general parameterless relation valid, in
time reversible systems, independently of the number of degrees of
freedom: the CH allows us to consider it as a manifestation of time
reversal symmetry.

\def\SEC{Reversible thermostats and Entropy Creation}
\section{Reversible thermostats and Entropy Creation}
\label{sec9}\iniz

Recalling that kinetic theory developed soon after the time average of
a mechanical quantity, namely kinetic energy, was understood to
have the meaning of absolute temperature, it is tempting to consider
quite important that, from the last three decades of research on
nonequilibrium statistical Mechanics, an interpretation emerged
of the physical meaning of the mechanical quantity $\s$ = phase space
contraction.

A system in contact with thermostats can generate entropy in the sense
that it can send amounts of heat into the thermostats thus increasing
their entropy by the ratio of the heat to the temperature, because the
thermostats must be considered in thermal equilibrium.

Furthermore if phase space contraction can be identified with a
physical quantity, accessible by means of calorimetric/thermometric
measurements, then the FT prediction becomes relevant and observable
and the CH can be subjected to tests, {\it independently on the
microscopic model that one may decide to assume}, which therefore
become possible also in real experiments.

It turns out that in very general thermostat models entropy production
rate can be identified with phase space contraction {\it up to a
``total time derivative''}: and since additive total time derivatives
(as we shall see) do not affect the asympotic distribution of time
averages, one can derive a FR for the entropy production (a quantity
accessible to measurement) from a FR for phase space contraction (a
quantity, in general, {\it not accessible} except in numerical
simulations, because it requires a precise model for the system, as a
rule not available).

As an example, of rather general nature, consider the following
one, obtained by imagining a system which is in contact with thermostats
that are ``external'' to it.  The particles of the system $\CC_0$ are
enclosed in a container, also called $\CC_0$, with elastic boundary
conditions surrounded by a few thermostats which consist of particles,
all of unit mass for simplicity, interacting with the system via short
range interactions, through a portion $\partial_i{\CC_0}$ of the surface
of ${\CC_0}$, and subject to the constraint that the total kinetic
energy of the $N_i$ particles in the $i$-th thermostat is
$K_i=\frac{1}2 \dot{\V X}_i^2=\frac32 N_i k_B T_i$.  A symbolic
illustration is in Fig.1.

\eqfig{110}{90}{}{fig.eps}{}

\0{\nota Fig.1: Particles in $\CC_0$ (``system particle'') interact
with the particles in the shaded regions (``thermostats particles'');
the latter are constrained to have a fixed total kinetic energy.}  \*

The equations of motion will be (all masses equal for simplicity)

\be \eqalign{
m\ddot{\V X}_0=&-\partial_{\V X_0}\Big( U_0(\V X_0)+\sum_{j>0}
W_{0,j}(\V X_{0},\V X_j)\Big)+\V E(\V X_0),
\cr
m\ddot{\V X}_i=&-\partial_{\V X_i}\Big( U_i(\V X_i)+
W_{0,i}(\V X_{0},\V X_i)\Big)-\a_i \dot{\V X}_i\hbox{\hglue1.1truecm}
\cr}\label{9.1}\ee
with $\a_i$ such that $K_i$ is a constant. Here $W_{0,i}$ is the
interaction potential between particles in $\CC_i$ and in $\CC_0$,
while $U_0,U_i$ are the internal energies of the particles in
$\CC_0,\CC_i$ respectively. We imagine that the energies $W_{0,j},U_j$
are due to {\it smooth} translation invariant pair
potentials; repulsion from the boundaries of the containers will be
elastic reflection. 

It is assumed, in Eq.(\ref{9.1}), that there is no direct interaction
between different thermostats: their particles interact directly only
with the ones in $\CC_0$.  Here $\V E({\V X}_0)$ denotes possibly
present external positional forces stirring the particles in
$\CC_0$. The contraints on the thermostats kinetic energies give

\be \a_i\equiv \frac{Q_i-\dot U_i}{3N_i k_B T_i}\qquad \otto\qquad
K_i\equiv const\defi\frac32 N_i k_B T_i\label{9.2}\ee
where $Q_i$ is the work per unit time that particles outside the thermostat
$\CC_i$ (hence in $\CC_0$) exercise on the particles in it, namely

\be Q_i\defi-\dot{\V
X}_i\cdot\partial_{{\V X}_{i}}W_{0,i}(\V X_{0},{\V X}_i)\label{9.3}\ee 
and it will be interpreted as the ``{\it amount of heat}'' $Q_i$ entering
the thermostat $\CC_i$ per unit time.

The main feature of the model is that the thermostats are
external to the system proper: this makes the model suitable for the
study of situations in which no dissipation occurs in the interior of
a system but it occurs only on the boundary.  

The {\it divergence} $-\s(\dot{\V X},\V X)$ of the equations of
motion, which gives the rate of contraction of volume elements around
$d\dot{\V X}d\V X$, does not vanish and can be computed in the model
in Fig.1; simple algebra yields, remarkably,

\be \eqalign{
\s(\V{{\dot X}},\V X)=&\,\e(\V{{\dot X}},\V X)+\dot R(\V X),\cr
\e(\V{{\dot X}}, \V X)=&\sum_{j>0} \frac{Q_j}{k_B T_j},
\qquad R(\V X)= \sum_{j>0} \frac{U_j}{k_B T_j}\cr}
\label{9.4}\ee
where $\e(\V{{\dot X}},\V X)$ can be interpreted as the {\it entropy
production rate}, because of the meaning of $Q_i$ in Eq.(\ref{9.3}).%
\footnote{\nota Eq.(\ref{9.4}) are correct up to $O(N^{-1})$ if
$N=\min N_j$ because the addends should contain also a factor
$(1-\frac1{3 N_j})$ to be exact: for simplicity $O(1/N)$ corrections
will be ignored here and in he following (their inclusion would imply
trivial changes without affecting the physical interpretation),
\cite{Ga06c}.\label{7}\label{footnote7}}

This is an interesting result because of its generality: it has
implications for the thermostated system considered in Fig.1, for
instance. It is remarkable that the quantity $p$ has a simple physical
interpretation: Eq.(\ref{9.1}) shows that the functions $\z_\s(p)$ and
$\z_\e(p)$ are {\it identical} because, since $R$ is bounded by our
assumption of smoothness, Eqs.
(\ref{9.2}) and (\ref{9.3}) imply 

\be \frac1\t \int_0^\t {\s(S_t(\dot{\V X},\V X))}dt
\equiv  \frac1\t \int_0^\t
{\e(S_t(\dot{\V X},\V X))}dt +\frac{R(\t)-R(0)}{\t},\label{9.5}\ee
so that

\be \s_+=\lim_{\t\to\infty}\frac1\t \int_0^\t \s(S_t(\dot{\V X},\V X))dt\equiv 
\lim_{\t\to\infty}\frac1\t \int_0^\t \e(S_t(\dot{\V X},\V X))dt=\e_+
\label{9.6}\ee
and the asmptotic distributions of 

\be p'=\frac1\t \int_0^\t 
\frac{\s(S_t(\dot{\V X},\V X))}{\s_+}dt, \qquad\hbox{and of}\qquad
p=\frac1\t \int_0^\t 
\frac{\e(S_t(\dot{\V X},\V X))}{\e_+}dt\label{9.7}\ee
{\it are the same}.

The Eq.(\ref{9.1}) are time reversible (with $I(\dot{\V X},\V
X)=(-\dot{\V X},\V X)$): then under the CH the large deviations rate
$\z(p)$ for the observable $\frac{\s}{\s_+}$ satisfies the ``{\it
fluctuation relation}'', Eq.(\ref{7.1}). It also follows that the
large deviations rate for $\frac{\e}{\e_+}$, identical to $\z(p)$,
satisfies it as well.

The point is that $\e$ is measurable by ``calorimetric and
thermometric measurements'', given its interpretation of entropy
increase of the thermostats. Therefore the CH can be subjected to
test or it can be used to ``predict'' the frequency of occurence of
unlikely fluctuations.
\*

\0{\it Comment:} This is a rather general example of thermostats
action, but it is just an example. For instance it can be generalized
further by imagining that the system is thermostatted in its
interior. A situation that arises naturally in the theory of electric
conduction. In the latter case the electrons move across the lattice
of the metal atoms and the lattice oscillations, {\it i.e.}  the
phonons, absorb or give energy. This can be modeled by adding a
``inner'' thermostat force $-\a_0\dot{\V x}_i$, acting on the
particles in $\CC_0$, which fixes the temperature of the electron
gas. Actualy a very similar model appeared in the early days of
Statistical Mechanics, in Drude's theory of electric conductivity,
\cite{Be64}. Other examples can be found in \cite{Ga06c}.

\def\SEC{Fluids}
\section{Fluids}
\label{sec10}\iniz

The attempt to put fluids and turbulence within the context provided
by the ideas exposed in the previous sections forces to consider cases
in which dissipation takes place irreversibly. This leads us to a few
conjectures and remarks. 

To bypass the obstacle due to the nonreversibility of the fluid
    equations which, therefore, seem quite far from the equations
    controlling the thermostated systems just considered, the
    following ``equivalence conjecture'', \cite{Ga02},  has been
    formulated. Consider the two equations for an incompressible flow
    with velocity field $\V u(\V x,t)$, $\BDpr\cdot\V u=0$, in
    periodic boundary condition for simplicity,

\be \eqalign{
&\dot{\V u}+\T {\V u}\cdot \T\partial\, {\V u}=\n \D\V u-\partial  p+ \V
g,\cr
&\dot{\V u}+\T {\V u}\cdot \T\partial {\V u}=\a(\V u) \D\V u-\partial  p+ \V
g,\cr}\label{10.1}\ee
where $\a(\V u)=\frac{\int \V u\cdot \V g\,d\V x}{\int (\partial \V u)^2
\,d\V x}$ is a ``Lagrange multiplier'' determined so that the total
energy $\EE\defi \int \V u^2\,d\V x$ is exactly constant.

Note that velocity reversal $I:\,\V u(\V x)\to -\V u(\V x)$
anticommutes, in the sense of Eq. (\ref{2.1}), with the time evolution
generated by the second equation (because $\a(I\V u)=-\a(\V u)$),
which means that ``fluid elements'' retrace their paths with opposite
velocity.

Introduce the ``local observables'' $F(\V u)$ as functions depending only
upon finitely many Fourier components of $\V u$, {\it i.e.}  on the ``large
scale'' properties of the velocity field $\V u$.
Then, {\it conjecture}, \cite{Ga97b}, the two equations should
have ``same large scale statistics'' in the limit $R\to+\infty$. If
$\m_\n$ and $\wt\m_\EE$ denote the respective SRB distributions of the
first and the second equations in Eq. (\ref{10.2}), by
{\it ``same statistics''} as $R\to\infty$ it is meant that 
\* 

\0(1) if the total energy $\EE$ of the initial datum $\V u(0)$ for the
second equation is chosen equal to the average $\media{\int \V
u^2\,d\V x}_{\m_\n}$ for the SRB distribution $\m_\n$ of the first equation,
then 
\\
(2) the two SRB distributions $\m_\n$ and $\wt\m_\EE$ are such
that, in the limit $R\to\infty$, the difference
$\media{F}_{\m_\n}-\media{F}_{{\wt\m}_\EE}\tende{R\to+\infty}0$.

\*
So far {\it only numerical tests} of the conjecture, in strongly cut off
$2$-dimensional equations, have been attempted (\cite{RS99}).
\*

An analogy with the termodynamic limit appears naturally: namely the
Reynolds number plays the role of the volume, locality of observables
becomes locality in $\V k$-space, and $\n,\EE$ play the role of
canonical temperature and microcanonical energy of the SRB
distributions of the two different equations in (\ref{10.1}),
respectively $\m_\n$ and $\wt\m_\EE$.  \*

The analogy suggests to question whether reversibility of the second
equation in Eq.(\ref{10.1}) can be detected. In fact to be able to see
for a large time a viscosity opposite to the value $\n$ would be very
unphysical and would be against the spirit of the conjecture.

If the CH is supposed to hold it is possible to use the FT, which is a
consequence of reversibility, to estimate the probability that, say,
the value of $\a$ equals $-\n$. For this purpose we have to first
determine the attracting set.

Assuming the K41, \cite{Ga02}, theory of turbulence the attracting set will
be taken to be the set of fields with Fourier components $\V
u_{\V k}=0$ unless $|\V k|\le R^{\frac34}$.

Then the expected identity $\media{\a}=\n$, between the average
 friction $\media\a$ in the second of Eq.(\ref{10.1}) and the viscosity
 $\n$ in the first, implies that the divergence of the evolution in the
 second of Eq.(\ref{10.1}) is in average

\be 
\s\sim \n \,\sum_{|\V k|\le R^{3/4}}2|\V k|^2\sim\,\n\,(\frac{2\p}L)^2
\frac{8\p}5 R^{15/4}\label{10.2}\ee
\*
By FT the SRB-probability to see, in motions following the second
equation in Eq. (\ref{10.2}), a ``{\it wrong}'' average friction $-\n$
for a time $\t$ is

\be  {\rm Prob}_{srb}\sim \exp{\big(-\t \n \frac{32\p^3}{5L^2} R^{\frac{15}{4}
}\big)}\,\defi\, e^{-g\t}\label{10.3}\ee
It can be estimated in the situation considered below for a flow in air:

\be \left\{\eqalign{
  \n=&  1.5\,10^{-2}\,\frac{cm^2}{sec},\quad v=10.\,\frac{cm}{sec}\,
\quad L=100.\,cm\cr
   R=&   6.67\,10^{4},\quad   g=3.66\,10^{14}\, sec^{-1}\cr
   P\defi& {\rm Prob}_{srb}=
    e^{-g\t}=e^{-3.66\,10^8},\qquad {\rm if}\quad 
\t=10^{-6}\cr}\right.\label{10.4}\ee
where the first line are data of an example of fluid motion and the
other two lines follow from Eq.(\ref{10.3}).  They show that, by FT,
viscosity can be $-\n$ during $10^{-6}s$ ({\it say}) with probability
$P$ as in Eq.(\ref{10.4}): unlikelyhood is similar in spirit to the
estimates about Poincar\'e's recurrences, \cite{Ga02}.  \*

\0(2) If we imagine that the particles are so many that the system can
be well described by a macroscopic equation, like for instance the NS
equation, then there will be two ways of computing the entropy
creation rate. The first would be the classic one described for
instance in \cite{DGM84}, and the second would simply be the
divergence of the microscopic equations of motion in the model of
Fig.1, under the assumption that the motion is closely described by
macroscopic equations for a fluid in local thermodynamic equilibrium,
like the NS equations. This can be correct in the limit in which space
and time are rescaled by $\e$ and $\e^2$ and the velocity field by
$\e$, and $\e$ is small. Since local equilibrium is supposed, it
will make sense to define a local entropy density $s(\V x)$ and a
total entropy of the fluid $S=\int s(\V x)\,d\V x$.

The evaluation of the expression for the entropy creation rate as a
divergence $\s$ of the microscopic equations of motion leads to,
\cite{Ga06}, a value $\media\e$ with average (over a microscopically
long time short with respect to the time scale of the fluid evolution)
related to the classical entropy creation rate in a NS fluid as

\be \eqalign{k_B \media\e=&k_B\e_{classic}+\dot S, \cr
k_B \e_{classic}=&\int_{\CC_0}\Big(\k\, 
\big(\frac{\V\partial T}{T}\big)^2
+\h\, \frac1T{\W{\Bt}'\cdot\W\partial \V u}\Big)\,d\V x\cr}\label{10.5}\ee
where $\W{\Bt}'$ is the tensor $(\partial_i u_j+\partial_j u_i)$ and
$\h$ is the dynamic viscosity, so that the two expressions differ by
the time derivative of an observable, which equals the total
equilibrium entropy of the fluid $S=\int s(\V x)\, d\V x$ where $s $
is the thermodynamical entropy density in the assumption of local
equilibrium; see comment on additive total derivatives preceding
Fig.1.

\def\SEC{Quantum Systems}
\section{Quantum Systems}
\label{sec11}\iniz

Recent experiments deal with properties on mesoscopic and atomic
scale. In such cases the quantum nature of the systems may not always
be neglected, paricularly at low temperature, and the question is
whether a fluctuation analysis parallel to the one just seen in the
classical case can be performed in studying quantum phenomena.

Thermostats have, usually, a macroscopic phenomenological nature: in a
way they should be regarded as classical macroscopic objects in which
no quantum phenomena occur.  Therefore it seems natural to model them
as such and define  their temperature as the average kinetic
energy of their constituent particles so that the question of how to define
it does not arise.

Consider the system in Fig.1 when the quantum nature of the particles
in $\CC_0$ cannot be neglected. Suppose for simplicity (see \cite{Ga07})
that the nonconservative force $\V E(\V X_0)$ acting on $\CC_0$
vanishes, {\it i.e.} consider the problem of heat flow through
$\CC_0$.  Let $H$ be the operator on $L_2(\CC_0^{3N_0})$, space of
symmetric or antisymmetric wave functions $\Ps(\V X_0)$,

\be H=
-\frac{\hbar^2}{2m}\D_{\V X_0}+ U_0(\V X_0)+\sum_{j>0}\big(U_{0j}(\V X_0,\V
X_j)+U_j(\V X_j)+K_j\big)\label{11.1}\ee
where $\D_{\V X_0}$ is the Laplacian, and note that its spectrum
consists of eigenvalues $E_n=E_n(\{\V X_j\}_{j>0})$, for $\V X_j$
fixed (because the system in $\CC_0$ has finite size).

A system--reservoirs model can be the {\it dynamical system} on the
space of the variables $\big(\Ps,(\{\V X_j\},$ $\{\V{{\dot
X}}_j\})_{j>0}\big)$ defined by the equations (where
$\media{\cdot}_\Ps\,=$ expectation in the state $\Ps$)

\be \eqalign{
-i\hbar {\dot\Ps(\V X_0)}=& \,(H\Ps)(\V X_0),\kern20mm{\rm and\ for}\
j>0\cr
\V{{\ddot X}}_j=&-\Big(\partial_j U_j(\V X_j)+
\media{\partial_j U_j(\V X_0,\V X_j)}_\Ps\Big)-\a_j \V{{\dot X}}_j\cr
\a_j\defi&\frac{\media{W_j}_\Ps-\dot U_j}{2 K_j}, \qquad
W_j\defi -\V{{\dot X}}_j\cdot \V\partial_j U_{0j}(\V X_0,\V
X_j)}\label{11.2} \ee
here the first equation is Schr\"odinger's equation, the second is an
equation of motion for the thermostats particles similar to the one in
Fig.1, (whose notation for the particles labels is adopted here
too). The model has no pretention of providing a physically correct
representation of the motions in the thermostats nor of the
interaction system thermostats, see comments at the end of this
section.

Evolution maintains the thermostats kinetic energies $K_j\equiv
\frac12\V{{\dot X}}_j^2$ exactly constant, so that they will be used
to define the thermostats temperatures $T_j$ via $K_j=\frac32 k_B T_j
N_j$, as in the classical case.

Let $\m_0(\{d\Ps\})$  be the {\it formal} measure on
$L_2(\CC_0^{3N_0})$ 

\be \Big(\prod_{\V X_0} d\Ps_r(\V X_0)\,d\Ps_i(\V X_0)
\Big)\,\d\Big(\int_{\CC_0} |\Ps(\V Y)|^2\, d\V Y-1\Big)
\label{11.3}\ee
with $\Ps_r,\Ps_i$ real and imaginary parts of $\Ps$.  The meaning of
(\ref{11.3}) can be understood by imagining to introduce an
orthonormal basis in the Hilbert space and to ``cut it off'' by
retaining a large but finite number $M$ of its elements, thus turning
the space into a high dimensional space $C^M$ (with $2M$ real
dimensions) in which $d\Ps=d\Ps_r(\V X_0)\,d\Ps_i(\V X_0)$ is simply
interpreted as the normalized euclidean volume in
$C^M$.

The formal phase space volume element $\m_0(\{d\Ps\})\times\n(d\V
X\,d\V{{\dot X}})$ with 

\be \n(d\V X\,d\V{{\dot X}})\defi\prod_{j>0} \Big(\d(\V{{\dot
X}}^2_j-3N_jk_B T_j)\,d\V X_j\,d\V{{\dot X}}_j\Big)
\label{11.4}\ee
is conserved, by the unitary property of the wave
functions evolution, just as in the classical case, {\it up
to the volume contraction in the thermostats}, \cite{Ga06c}. 

If $Q_j\defi\media{W_j}_\Ps$ and $R$ is as in Eq.(\ref{9.4}), then the
contraction rate $\s$ of the volume element in Eq.(\ref{11.4}) can be
computed and is (again) given by Eq.(\ref{9.4}) with $\e$, that will
be called {\it entropy production rate}: 
setting $R(\V X)\defi \sum_{j>0} \frac{U_j(\V X_j)}{k_B T_j}$, it is

\be\s(\Ps,\V{{\dot X}},\V X)=\,\e(\Ps,\V{{\dot X}},\V X)+\dot R(\V X),\qquad
\e(\Ps,\V{{\dot X}}, \V X)=\sum_{j>0} \frac{Q_j}{k_B T_j},
\label{11.5}\ee

In general solutions of Eq.(\ref{11.2}) {\it will not be quasi periodic} and
the Chaotic Hypothesis, \cite{GC95b,Ga00,Ga07}, can be assumed: if so the
dynamics should select an SRB distribution $\m$. The
distribution $\m$ will give the statistical properties of the
stationary states reached starting the motion in a thermostat
configuration $(\V X_j,\V{{\dot X}}_j)_{j>0}$, randomly chosen with
``uniform distribution'' $\n$ on the spheres $m\V{{\dot X}}_j^2=3N_jk_B
T_j$ and in a random eigenstate of $H$. The distribution $\m$, if
existing and unique, could be named the {\it SRB distribution}
corresponding to the chaotic motions of Eq.(\ref{11.2}).

In the case of a system {\it interacting with a single thermostat} at
temperature $T_1$ the latter distribution should be equivalent to the
canonical distribution, up to boundary terms.

Hence an important consistency check, for proposing Eq.(\ref{11.2}) as
a model of a thermostated quantum system, is that there should exist
at least one stationary distribution equivalent to the canonical
distribution at the appropriate temperature $T_1$ associated with the
(constant) kinetic energy of the thermostat: $K_1=\frac32 k_B
T_1\,N_1$.  In the corresponding classical case this is an established
result, \cite{EM90,Ga00,Ga06c}.

A natural candidate for a stationary distribution could be to
attribute a probability proportional to $d\Ps\,d\V X_1\,d \dot{\V
X}_1$ times

\be 
\sum_{n=1}^\infty e^{-\b_1 E_n}\d(\Ps-\Ps_n(\V
X_1)\,e^{i\f_n})\,{d\f_n}\,\d(\dot{\V X}_1^2-2K_1)\label{11.6}\ee
where $\b_1=1/k_B T_1$, $\Ps$ are wave functions for the system in
$\CC_0$, ${\dot {\V X}_1, \V X_1}$ are positions and velocities of
the thermostat particles and $\f_n\in [0,2\p]$ is a phase, $E_n=E_n(\V
X_1)$ is the $n$-th level of $H(\V X_1)$, with $\Ps_n(\V X_1)$ the
corresponding eigenfunction. The average value of an observable $O$
for the system in $\CC_0$ in the distribution $\m$ in (\ref{11.6}) would
be

\be \media{O}_\m=Z^{-1}\int {\rm Tr}\, (e^{-\b H(\V X_1)} O)\,\d(\dot{\V
X}_1^2-2K_1)d\V X_1\,d \dot{\V X}_1\label{11.7}\ee
where $Z$ is the integral in (\ref{11.7}) with $1$ replacing $O$,
(normalization factor).  Here one recognizes that $\m$ attributes to
observables the average values corresponding to a Gibbs state at
temperature $T_1$ with a random boundary condition $\V X_1$.

However Eq.(\ref{11.6}) {\it is not invariant} under the evolution
Eq.(\ref{11.2}) and it seems difficult to exhibit explicitly an
invariant distribution. Therefore one can say that the SRB
distribution for the evolution in (\ref{11.2}) is equivalent to the
Gibbs distribution at temperature $T_1$ only as a conjecture.

Nevertheless it is interesting to remark that under the {\it adiabatic
approximation} the eigenstates of the Hamiltonian at time $0$ evolve
by simply following the variations of the Hamiltonian $H(\V X(t))$ due
to the motion of the thermostats particles, without changing quantum
numbers (rather than evolving following the Schr\"odinger equation and
becoming, therefore, different from the eigenfunctions of $H(\V
X(t))$).

In the adiabatic limit in which the classical motion of the
thermostat particles takes place on a time scale much slower than the
quantum evolution of the system the distribution (\ref{11.6}) {\it is
invariant}. 

This can be checked by first order perturbation analysis
which shows that, to first order in $t$, the variation of the energy
levels (supposed non degenerate) is compensated by the phase space
contraction in the thermostat, \cite{Ga07}. 
Under time evolution, $\V X_1$ changes, at time $t>0$, into $\V X_1+t
\V{{\dot X}}_1+O(t^2)$ and, assuming non degeneracy, the eigenvalue
$E_n(\V X_1)$ changes, by perturbation analysis, into $E_n+t \,
e_n+O(t^2)$ with

\be e_n\defi t\media{\V{{\dot X}}_1\cdot\V\partial_{\V X_1}
U_{01}}_{\Ps_n}+t \V{{\dot X}}_1\cdot\V\partial_{\V X_1}
U_{1}=-t\,(\media{W_1}_{\Ps_n}+\dot R_1)=-\frac1{\b_1}\a_1.\label{11.8}
\ee
Hence the Gibbs factor changes by $e^{-\b t e_n}$ and at the same time
phase space contracts by $e^{t \frac{3 N_1 e_n}{2K_1}}$, as it follows
from the expression of the divergence in Eq.(\ref{11.5}). {\it
Therefore if $\b$ is chosen such that $\b=(k_B T_1)^{-1}$ the state
with distribution Eq.(\ref{11.6}) is stationary}, (recall that for
simplicity $O(1/N)$, see footnote${}^{\ref{footnote7}}$ on
p.\pageref{7} 
is neglected). This shows that, {\it in the adiabatic approximation},
interaction with only one thermostat at temperature $T_1$ admits at
least one stationary state. The latter is, by construction, a Gibbs
state of thermodynamic equilibrium with a special kind (random $\V
X_1,\V{{\dot X}}_1$) of boundary condition and temperature $T_1$.  \*

\0{\it Remarks:}
(1) The interest of the example is to show that even in quantum
    systems the chaotic hypothesis makes sense and the intepretation
    of the phase space contraction in terms of entropy production
    remains unchanged. In general, under the chaotic hypothesis, the
    SRB distribution of (\ref{11.2}) (which in presence of forcing, or
    of more than one thermostat is certainly quite not trivial, as in
    the classical Mechanics cases) will satisfy the fluctuation
    relation because the fluctuation theorem only depends on
    reversibility: so the model (\ref{11.2}) might be suitable (given
    its chaoticity) to simulate the steady states of a quantum system
    in contact with thermostats.

\0(2) It is certainly unsatisfactory that a stationary distribution
    cannot be explicitly exhibited for the single thermostat case (unless
    the adiabatic approximation is invoked). However, according to the
    proposed extension of the CH, the model does
    have a stationary distribution which should be equivalent (in the sense
    of ensembles equivalence) to a Gibbs distribution at the same
    temperature. 

\0(3) The non quantum nature of the thermostat considered here and the
    specific choice of the interaction term between system and
    thermostats should not be important: the very notion of thermostat
    for a quantum system is not at all well defined and it is natural
    to think that in the end a thermostat is realized by interaction
    with a reservoir where quantum effects are not
    important. Therefore what the analysis really suggests is
    that in experiments in which really microscopic systems are
    studied the heat exchanges of the system with the external world
    should fulfill a FR.

\0(4) The conjecture can probably be tested with present day
    technology. If verified it could be used to develop a
    ``Fluctuation Thermometer'' to perform temperature measurements
    which are {\it device independent} in the same sense in which the gas
    thermometers are device independent ({\it i.e.} do not require
    ``calibration'' of a scale and ``comparison''
    procedures).
\\
Consider a system in a stationary state, and imagine inducing small
    currents and measuring the average heat output rate $Q_+$ and the
    fluctuations in the finite time average heat output rate,
    generated by inducing small currents, {\it i.e.}  fluctuations of
    $p=\frac1\t\int_0^\t \frac{Q(t)}{Q_+}dt$ obtaining the rate function of
    $\z(p)$.
\\
Then it becomes possible to read from the slope of $\z(p)-\z(-p)$,
    equal to $\frac{Q_+}{k_B T}$ by the FR, directly the inverse
    temperature that the thermostat in contact with the system has:
    this could be useful particularly in very small systems (classical
    or quantum). The idea is inspired by a similar earlier proposal
    for using fluctuation measurements to define temperature in spin
    glasses, \cite{CKP97}, \cite[p.216]{CR03}.

\section{Experiments ?}
\def\SEC{Experiments ?}
\label{sec12}\iniz

The (partial) test of the chaotic hypothesis via its implication on
large fluctuations probabilities ({\it i.e.}  the fluctuation relation) is
quite difficult. The main reason is that if the forcing is small the
relation degenerates (because $\e_+\to0$) and it can be shown,
\cite{Ga96a}, that to lowest nontrivial order in the size of the forcing it
reduces to the Green-Kubo formula, which is (believed to be) well
established so that the fluctuation relation will not be significant,
being ``true for other reasons'', \cite{DGM84}. See Sec.3.

Hence one has to consider large forcing. However, under large forcing,
large fluctuations of $p$ become very rare, hence their statistics is
difficult to observe. Furthermore the statistics seems to remain
Gaussian for $p$, in a region around $p=1$ where the data can be
considered reliably unbiased (see below), and until rather large
values of the forcing field or values of $|p-1|$ large compared to the
root mean square deviation $\frac{D}{\sqrt\t}={\langle
(p-1)^2\rangle^{1/2}}$ are reached. Hence
$\z(p)=-\frac1{2D^2} (p-1)^2$ and linearity in $p$ of $\z(p)-\z(-p)$
is trivial. {\it Nevertheless}, in this regime, it follows that
$\frac2{D^2}=\s_+$ which is a nontrivial relation and therefore a
simple test of the fluctuation relation.

The FR was empirically observed first in such a situation in
\cite{ECM93}, in a simulation, and the first dedicated tests, after
recognizing its link with the CH, were still performed in a Gaussian
regime, so that they were really only tests of $\frac2{D^2}=\s_+$ and of
the Gaussian nature of the observed fluctuations.

Of course in simulations the forcing can be pushed to ``arbitrarily
large'' values so that the fluctuation relation can, in principle, be
tested in a regime in which $\z(p)$ is sensibly non Gaussian, see
\cite{LLP98}. But far more interesting will be cases in which the
distribution $\z(p)$ is sensibly not Gaussian and which deal with
laboratory experiments rather than simulations. Skepticism towards
the CH is mainly based on the supposed non measurability of the
function $\z(p)$ in the large deviation domain ({\it i.e.} $|p-1|\gg
\sqrt{\langle(p-1)^2\rangle}$).

In experimental tests several other matters are worrysome,
among which: 
\*

\0(a) is reversibility realized? This is a rather stringent and
difficult point to understand on a case by case basis, because
irreversibility creeps in, inevitably, in dissipative phenomena.

\0(b) is it allowed to consider $R$, {\it i.e.}  the 
``entropy production remainder'' in (\ref{9.3}), bounded? if not there
will be corrections to FR to study (which in some cases,
\cite{CV03a,BGGZ05}, can be studied quite in detail).

\0(c) does one introduce any bias in the attempts to see statistically
large deviations? for instance in trying to take $\t$ large one may be
forced to look at a restricted class of motions, typically the ones
that remain observable for so long a time. It is easy to imagine that
motions observed by optical means, for instance, will remain within
the field of the camera only for a characteristic time $\t_0$ so that
any statistics on motions that are observed for times $\t>\t_0$ will
be biased (for it would deal with untypical events).

\0(d) chaotic motions may occur under influence of stochastic
perturbations, so that extensions of FT to stochastic systems may need
to be considered. This is not really a problem because a random
perturbation can be imagined as generated by coupling of the system to
another dynamical system (which, for instance, in simulations would be
the random number generator from which the noise is drawn),
nevertheless it demands careful analysis, \cite{BGG07}.

\0(e) Nonconvex shape of $\z(p)$, at $|p-1|$ beyond the root mean square
 deviation, see Fig.3, is seen often, possibly always, in the
 experiments that have been attempted to study large
 deviations. Therefore the interpretation of the nonconvexity, via
 well understood corrections to FR, seems to be a forced path towards
 a full test of the FR, beyond the Gaussian regime, \cite{BGGZ05}.  

\*
\eqfig{0}{125}{\ins{-100}{80}{$10\t_C$}}{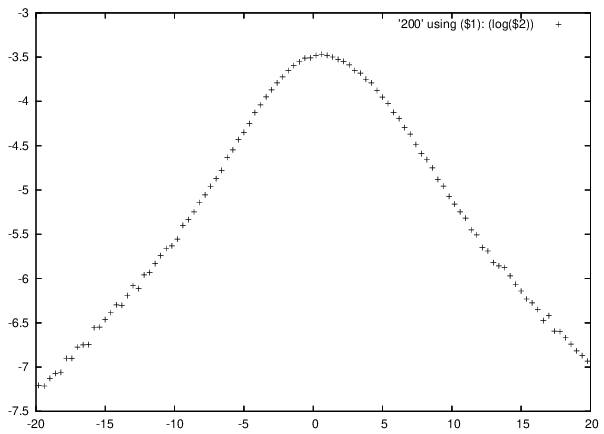}{}

\0{\nota Fig.3: An histogram of ${\st \log} P_{\t}(p)$, taken from the
 data of \cite{BCG07} at time $\t=10\t_C=200$ms: it shows the rather
 typical nonconvexity for $|p-\st1|\sim8$ which is of the order of
 standard deviation.}  
\* 

All the above questions arise in the recent experiment by
Bandi-Cres\-sman-Goldburg, \cite{BCG07}. It encounters all the related
difficulties and to some extent provides the first evidence for the FR
(hence the CH) in a system in which the predictions of the FR are not
the result of a theoretical model which can be solved exactly. The
interpretation of the results is difficult and further investigations
are under way.

The experiment outcome is not incompatible with FR and, in any event,
it proves that good statistics can be obtained for fluctuations that
extend quite far beyond the root mean square deviation of $p-1$: an
asset of the results in view of more refined experiments.

A very promising field for experimental tests of the CH and the FR 
is granular materials: in granular materials collisions are not
elastic, nevertheless an experiment is proposed in \cite{BGGZ06}.
See comment (6) in Sec.13 and comment (4) to Eq. (\ref{11.8}) 
for other hints at possible experiments and applications.

\section{Comments}
\def\SEC{Comments}
\label{sec13}\iniz

(1) In the context of the finite thermostats approach, besides systems of
particles subject to deterministic evolution, stochastically evolving
systems can be considered and the FT can be extended to cover the new
situations, \cite{Ku98,LS99,Ma99,CDG07,BGG07}.
\\
(2) Alternative quantum models have also been considered in the
literature, \cite{Ku00} (stochastic Langevin thermostats), or infinite 
thermostats (free and interacting, and possibly with further noise
sources) \cite{FV63,JP02b,HI05,APJP06,JOP07}.
\\
(3) Many simulations have been performed, starting with the
experiment which showed data that inspired the FT, \cite{ECM93}, and
continuing after the proof of FT and the formulation of the CH, {\it e.g.}
\cite{BGG97}: a few had the purpose of testing the Fr in a nongaussian
regime for the fluctuations of the variable $p$, \cite{LLP98}. In some
cases the results had to be examined closely to understand what was
considered at discrepancy with the FT, \cite{BGGZ05}, (and was not). 
\\
(4) The physical relevance of the particular quantum thermostat model
remains an open question and essentially depends on the conjecture
that the (unknown) SRB distribution for the model in the single
thermostat case is equivalent to the Gibbs distribution at the same
temperature (a property valid in the corresponding classical cases).
Hence the main interest of the model is that it shows that a FR is in
principle possible in finite thermostated quantum systems in
stationary state.
\\
(5) Few experiments have so far been performed (besides numerical
    simulations) to investigate CH and FT: extensions to randomly
    forced systems are possible, \cite{Ku98,LS99,Ma99}, and can be
    applied to systems that can be studied in laboratory,
    \cite{CDGS04,BCG07}: the first experiment designed to test the FR
    in a laboratory experiment is the recent work \cite{BCG07}. The
    results are consistent with the FR and indicate a promising
    direction of research.
\\
(6) An interesting consequence of the FT is that

\be \media{e^{-\D S/k_B}}_{srb}\defi 
\media{ e^{-\int_0^\t \sum_{j>0}\frac{Q_j(t)}{k_B T_j}} dt}_{srb}= O(1)
\label{13.1}\ee
    in the sense that the logarithms of both sides divided by $\t$
    agree in the limit $\t\to\infty$ ({\it i.e.}
    $\lim_{\t\to+\infty}\frac1\t\log \media{e^{\D S/k_B}}=0$) with
    corrections of order $O(\frac1\t)$. This has been pointed out by
    Bonetto, see \cite{Ga00}, and could have applications in the same
    biophysics contexts in which the work theorems, \cite{Ja97,Ja99},
    have been applied: for instance one could study stationary heat
    exchanges is systems out of equilibrium (rather than measure free
    energy differences between equilibrium states at the same
    temperature as in \cite{Ja97,Ja99}). The boundedness of the
    l.h.s. of Eq. (\ref{13.1}) implied by (\ref{13.1}) can be used to
    test whether some heat emissions have gone undetected (which would
    imply that the l.h.s. of Eq.(\ref{13.1}) tends to $0$, rather than
    staying of $O(1)$). This is particularly relevant as in biophysics
    one often studies systems in stationary states while actively busy
    at exchanging heat with the sourroundings.
\\
\0(7) Another property, which is not as well known as it deserves, is
    that for hyperbolic systems, and by the Chaotic Hypothesis of
    Sec. 2, virtually for all chaotic evolutions, it is possible to
    develop a rigorous theory of coarse graining, \cite{BG97,Ga06b}.
    It leads to interpreting the SRB distributions as uniform
    distributions on the attractor; hence to a variational principle
    and to the existence of a Lyapunov function describing the
    approach to the stationary state, {\it i.e.}  giving a measure of
    the distance from it, \cite{Ga01,Ga06}.
\\
However it also seems to lead to the conclusion that {\it entropy} of
    a stationary state {\it cannot be defined} if one requires that it
    should have properties closely analogous to the equilibrium
    entropy. For instance once coarse graining has been properly
    introduced, it is tempting to define the entropy of a stationary
    state as $k_B$ times the logarithm of the number of ``microcells''
    into which the attractor is decomposed, see Appendix A1,A2. 
\\
This quantity can be used as a Lyapunov function, see \cite{Ga06}, but
it depends on the size of the microcells in a nontrivial way: changing
their size, the variation of the so defined entropy does not change by
an additive constant depending only on the scale of the coarse
graining ({\it at difference with respect to the equilibrium case}),
but by a quantity that depends also on the control parameters ({\it
e.g.}  temperature, volume {\it etc.\ }), \cite{Ga01}.
\\
Given the interest of coarse graining, in Appendix A1 mathematical
details about it are discussed in the context of the SRB distribution
and CH; and a physical interpretation is presented in Appendix A2;
hopefully they will also clarify the physical meaning of the two.
\\
(8) Finally it is often said that the FR should hold {\it always} or,
if not, it is incorrect. In this respect it has to be stressed that
the key assumption is the CH, which implies the FR {\it exactly} in
time reversible situations. However it is clear that CH is an
idealization and the correct attitude is to interpret deviations from
FR in terms of corrections to the CH.  For instance: \*

\0CH implies exponential decay of time correlations. But in
some cases there are physical reasons for long range time
correlations.
\\
Or the CH implies that observables have values in a
finite range. But there are cases in which phase space is not bounded
and observables can take unbounded values (or such for practical
purposes).
\\
Time reversal is necessary. But there are cases in which it
is violated.
\\
The pdf of $p$ should be log-convex: but it is seldom so.
\* 
What is interesting is that it appears that starting from CH and
examining the features responsible for its violations it may be
possible to compute even quantitatively the corrections to
FR. Examples of such corrections already exist,
\cite{CV03a,BGGZ05,Za06}. It would be interesting to have a
concrete experiment, designed to test FR and try to understand the
observed deviations; the BCG experiment in Sec.\ref{sec12} offers, if
further developed, the possibility of simple tests making use the
existing experimental apparatus and of the observations that it has
proved to be accessible.

\* \0{\bf Acknowledgements:} I am grateful to M. Bandi, A. Giuliani,
W. Goldburg and F. Zamponi for countless comments and suggestions and
to M. Bandi, W. Goldburg for providing their data, partially reported
in Fig.3. Partially supported also by Institut des Hautes Etudes 
Scientifiques, by Institut Henri Poincar\'e and by Rutgers University. \*

\section{A1: Coarse Graining, SRB and $1D$ Ising Models}
\def\SEC{A1: Coarse Grain, SRB and Ising Lattice}\label{secA1}\iniz

In equilibrium phase space volume is conserved and it is natural to
imagine it divided into tiny ``cells'', in which all observables of
interest are constant. The equilibrium distribution can be constructed
simply by imagining to have divided phase space $\Si$ (``energy
surface'') into cells of equal Liouville volume, small enough so that
every interesting physical observable $F$ is constant in each
cell. Then the dynamics is a cyclic permutation of the cells ({\it
ergodic hypothesis}) so that the stationary distribution is just the
volume distribution.

In a way, this is an ``accident'', based on what appears to be a
fundamentally incorrect premise, which leads to various difficulties
as it is often considered in the context of attempts to put on firm
grounds the notion of a ``coarse grained'' description of the
dynamics. Confusion is also added by the simulations: the latter are
sometimes interpreted as {\it de facto} coarse grained
descriptions. It seems, however, essential to distinguish between
coarse graining and representation of the dynamics as a permutation of
small but finite cells.

{\it Undoubtedly} dynamics can be represented by a permutation of small
phase space volumes, as any simulation program effectively does. But it
is also clear that the cells used in the simulations are {\it far too
small} ({\it i.e.}  of the size determined by the computer resolution,
typically of double precision reals) to be identified with the coarse
cells employed in phenomenological studies of statistical
Mechanics.

On the other hand if coarse grain cells are introduced which are not
as tiny as needed in simulations the {\it dynamics will deform}
them to an extent that after a short time it will no longer be
possible to identify which cell has become which other cell! And this
applies even to equilibrium states.

In this respect it looks as an accident the fact that, nevertheless, at least
in equilibrium a coarse grained representation of time evolution
appears possible. And easily so, with small cells subject to the only
condition of having equal volume; but the huge amount of literature on
attempts at establishing a theory of coarse graining did not lead to a
precise notion, nor to any agreement between different proposals.

Under the CH systems are hyperbolic and a precise analysis of coarse
graining seems doable, see \cite{Ga01,Ga95a} and \cite{Ga04b}.  The key is
that it is possible to distinguish between ``microcells'', so tiny
that evolution is well approximated by a permutation on them, and
``cells'' which are still so small that the (few) interesting
observables have constant value on them. The latter cells can be
identified with ``coarse grain cells''; yet they are very large
compared to the microcells and time evolution {\it cannot} be
represented as their permutation. {\it Neither in equilibrium nor out of
equilibrium.}

That SRB distribution cannot be considered a permutation of naively
defined coarse cells {\it seems} to be well known and to have been
considered a drawback of the SRB distributions: it partly accounts for
the skepticism that often, still now, accompanies them.  \*

{\it The point that will be made, {\rm see the review \cite{Ga04b}},
is that hyperbolicity provides us with a natural definition of coarse
grained cells. At the same time it tells us which is the weight to be
given to each cell which, in turn, implies that each cell can be
imagined containing many ``microcells'' whose evolution is a simple
permutation of them {\rm(just as in numerical simulations)}.}  \*

In this appendix we consider for simplicity discrete time systems: in
this case hyperbolic systems are described by a smooth map $S$,
transitive and smoothly invertible, with the property that every phase
space point $x$ is a ``saddle point''. Out of $x$ emerge the stable and the
unstable manifolds $W^s(x),W^u(x)$ of complementary dimension.  The
expansion and contraction that take place near every point $x$ can be
captured by the matrices $\partial S_u(x)$, $\partial S_s(x)$ obtained
by restricting the matrix ({\it Jacobian matrix}) $\partial S(x)$, of
the derivatives of $S$, to its action on the vectors tangent to the
unstable and stable manifolds through $x$: the evolution $S$ maps
$W^u(x)$,$W^s(x)$ to $W^u(Sx),W^s(Sx)$, respectively, and its
derivative ({\it i.e.}  its linearization) maps tangent vectors at $x$
into tangent vectors at $Sx$.
 
A quantitative expression of the expansion and contraction is given by
the ``local expansion'' or ``local contraction'' rates defined by

\be 
\L^u_1(x)\defi\log|\det (\partial S)_u(x)|,\qquad
\L^s_1(x)\defi-\log|\det (\partial S)_s(x)|.\label{14.1}
\ee
Since time is now discrete, phase space contraction is now defined as
$\s(x)=-\log|\det(\partial S)|$ and related to $\L^u_1(x),\L^s_1(x)$ by

\be \s(x)=-\L_1^u(x)+\L_1^s(x)-\log\frac{\sin\d(Sx)}{\sin\d(x)},\label{14.2}\ee
where $\d(x)$ is the angle (in the metric chosen in phase space)
between $W^s(x),W^u(x)$ (which is bounded away from $0$ and $\p$ by
the smoothness of the hyperbolic evolution $S$).

This suggests to imagine constructing a partition $\PP$ of phase space into
closed regions ${\cal P}=(P_1,\ldots,P_m)$ with pairwise disjoint
interiors, each of which is a ``rectangle'' defined as follows.

The rectangle $P_i$, see the following Fig.5 for a visual guide,
 has a center $\k_i$ out of which emerge portions $C\subset
W^s(\k_i),D\subset W^u(\k_i)$ of its stable and unstable manifolds,
small compared to their curvature, which form the ``axes'' of $P_i$,
see Fig.5.  The set $P_i$, then, consists of the points $x$ obtained
by taking a point $p$ in the axis $D$ and a point $q$ in the axis $C$
and setting $x{\buildrel c\over =}W^s(p)\cap W^u(q)$, just as in an
ordinary rectangle a point is determined by the intersection of the
lines through any two points on the axes and perpendicular to them,
see Fig.5. The symbol ${\buildrel c\over =}$ means that $x$ is the
point closest to $p$ and to $q$ along paths in $W^s(p)$ and,
respectively $W^u(q)$.\footnote{\nota This proviso is needed because
  often, and certainly in transitive hyperbolic maps, the full
manifolds $W^s(p),W^u(q)$ are dense in phase space and intersect
infinitely many times, \cite{Si72,Si77}.}

Note that in a rectangle {\it anyone} of its points $\Bk$ could be the
center in the above sense with a proper choice of $C,D$, so that
$\k_i$ does not play a special role and essentially serves as a label
identifying the rectangle. In dimension higher than $2$ the rectangles
may (and will) have rather rough (non differentiable) boundaries,
\cite{Bo78}.

\eqfig{130}{120}{
\ins{80}{85}{$C$}
\ins{74}{39}{$D$}
\ins{58}{69}{$\k$}
\ins{120}{96}{$W^s_\g(x)$}
\ins{115}{27}{$W^u_\g(x)$}
\ins{22}{14}{$P$}}
{fig5.eps}{}

\0{\nota Fig.5: A rectangle $P$ with a pair of axes $C,D$
crossing at the corresponding center $\k$.}
\*

It is a key property of hyperbolicity (hence of systems for which the
CH can be assumed) that the partition ${\cal P}$ can be built to enjoy
of a very special property.  

Consider the sequence, {\it history of
$x$}, $\Bx(x)\defi\{\x_i\}_{i=-\infty}^\infty$ of symbols telling into
which of the sets of ${\cal P}$ the point $S^ix$ is, {\it i.e.}
where $x$ is found at time $i$, or $S^ix\in P_{\x_i}$. This is
unambiguous aside from the zero volume set $\cal B$ of the points that
in their evolution fall on the common boundary of two $P_\x$'s.  

Define the matrix $Q$ to be $Q_{\x,\x'}=0$, unless there is an
interior point in $P_{\x}$ whose image is in the interior of
$P_{\x'}$: and in the latter case set $Q_{\x,\x'}=1$. Then the history
of a point $x$, which in its evolution does not visit a boundary common
to two $P_\x$'s, must be a sequence $\Bx$ verifying the property,
called {\it compatibility}, that, $Q_{\x_k,\x_{k+1}}=1$ for all times
$k$.

The matrix $Q$ tells us which sets $P_{\x'}$ can be reached from
points in $P_\x$ in one time step. Then transitive hyperbolic maps
admit a partition (in fact infinitely many) of phase space into
rectangles ${\cal P}=(P_1,\ldots,P_m)$, so that \*

\0(1) if $\Bx$ is a compatible sequence then there is a point $x$
such that $S^k x\in P_{\x_k}$, see (for instance) Ch. 9 in
\cite{Ga00}, (``{\it compatibility}''). The points $x$ outside
the exceptional set $\cal B$ (of zero volume) determine uniquely the
corresponding sequence $\Bx$.

\0(2) the diameter of the set of points
$E(\x_{-\frac12T},\ldots,\x_{\frac12T})$ consisting of all points
which between time $-\frac12T$ and $\frac12T$ visit, in their
evolution, the sets $P_{\x_i}$ is bounded above by $c\, e^{-c' T}$ for
some $c,c'>0$ ({\it i.e.}  the code $\V\x\to x$ determines $x$ ``{\it
with exponential precision}'').

\0(3) there is a power $k$ of $Q$ such that $Q^k_{\x\x'}>0$ for all
$\x,\x'$ (``{\it transitivity}'').
\*

Hence points $x$ can be identified with sequences of symbols $\V\x$
verifying the compatibility property and the sequences of symbols
determine, with exponential rapidity, the point $x$ which they
represent. 

The partitions ${\cal P}$ are called {\it Markov
partitions}. Existence of ${\cal P}$ is nontrivial and rests on the
chaoticity of motions: because the compatibility of all successive pairs
implies that the full sequence is actually the history of a point (a
clearly false statement for general partitions).\footnote{\nota The
Markovian property has a geometrical meaning: imagine each $P_i$ as
the ``stack'' of the connected unstable manifolds portions $\d(x)$,
intersections of $P_i$ with the unstable manifolds of its points $x$,
which will be called unstable ``layers'' in $P_i$. Then if
$Q_{i,j}=1$, the expanding layers in each $P_i$ expand under the
action of $S$ and their images {\it fully cover} the layers of $P_j$
which they touch.  Formally let $P_i\in\PP$ and $x\in P_i$,
$\d(x){\buildrel c\over=} P_i\cap W_u(x)$: the if $Q_{i,j}=1$, {\it i.e.}
if $SP_i$ visits $P_j$, it is $\d(Sx)\subset S\d(x)$.\label{footnote9}}

If the map $S$ has a time reversal symmetry $I$ ({\it i.e} a smooth
involution $I$, such that $IS=S^{-1}I$, see Eq.(\ref{2.1})) the
partition $\PP$ can be so built that $I\PP=\PP$, hence $I
P_i=P_{I(i)}$ for some $I(i)$. This is done simply by replacing $\PP$
by the finer partition whose elements are $P_i\cap IP_j$, because if
$\PP,\PP_1$ and $\PP_2$ are Markovian partitions also the partition
$I\PP$ is such, as well as the partition $\PP_1\vee\PP_2$ formed by
intersecting all pairs $P\in \PP_1$, $P'\in\PP_2$ (this is best seen
from the geometric interpretation in footonote$^{\ref{footnote9}}$ and
from the time reversal property that $I W_u(x)=W_s(Ix)$).

A Markov partition such that $I\PP=\PP$ is called ``reversible'' and
histories on it have the simple property that
$(\Bx(Ix))_i=(\Bx(x))_{-I(i)}$.

Markov partitions, when existing, allow us to think of the phase space
points as the configurations of a ``$1$-dimensional spin system'',
{\it i.e.}  as sequences of finitely many symbols
$\x\in\{1,2,\ldots,m\}$ subject to the ``hard core'' constraint that
$Q_{\x_i,\x_{i+1}}=1$. Hence probability distributions on phase space
which give $0$ probability to the boundaries of the elements of the
Markov partitions (where history may be ambiguous) can be regarded as
stochastic processes on the configurations of a $1$-dimensional Ising
model (with finite spin $m$), and functions on phase space can be
regarded as functions on the space of compatible
sequences.\footnote{\nota
It is worth also stressing that the ambiguity of the
histories for the points which visit the boundaries of the sets of a Markovian
partition is very familiar in the decimal representation of
coordinates: it corresponds to the ambiguity in representing a decimal
number as ending in infinitely many $0$'s or in infinitely many $9$'s.}

The remarkable discovery, see reviews in \cite{Si72,Si77}, is that the
SRB distribution not only can be regarded as a stochastic processes,
but it {\it is a short range Gibbs distribution} if considered as a
probability on the space of the compatible symbolic sequences $\Bx$ on
$\PP$, and with a potential function $A(\Bx)=-\L_1^u(x( \Bx))$,
see below and \cite{GBG04}.

The sequences $\Bx$ are therefore much more natural, given the
dynamics $S$, than the sequence of decimal digits that are normally
used to identify the points $x$ via their cartesian
coordinates.\footnote{\nota If the phase space points are considered
as sequences $\Bx$ then the dynamics becomes a ``trivial'' left shift
of histories. This happens always in symbolic dynamics, but in general
it is of little interest unless compatibility can be decided by a
``hard core condition'' involving only nearest neighbors (in general
compatibility is a global condition involving all symbols, {\it i.e.}
as a hard core it is one with infinite range). {\it Furthermore} also
the statistics of the motion becomes very well understood, because
short range $1D$ Gibbs distributions are elementary and well
understood.}  \*

\0{\bf Definition:} {\it ({\rm Coarse graining}) Given a Markovian
  partition $\PP$ let $\PP^T$ be the finer partition of phase space
  into sets of the form

\be E_{\Bx}=E_{\x_{-T/2},\ldots,\x_{T/2}}\defi\bigcap_{-T/2}^{T/2} S^k
P_{\x_k}.
\label{14.3}\ee 
The sets $E_\Bx$ will be called ``elements of a description of the
microscopic states {\rm coarse grained to scale $\g$}'' if $\g$ is the
largest linear dimension of the nonempty sets $E_{\Bx}$. The elements
$E_{\Bx}$ of the ``coarse grained partition $\PP^T$ of phase space''
are labeled by a finite string
\be \Bx\,=\, (\x_{-T/2},\ldots,\x_{T/2})\label{14.4}\ee
with $\x_i=1,\ldots,m$ and $Q_{\x_i,\x_{i+1}}=1$.}
\*

Define the {\it forward} and {\it backward} expansion and contraction
rates as

\be U^{T/2}_{u,\pm}(x)=\sum_{j=0}^{\pm T/2} \L_1^u(S^jx),\qquad
U^{T/2}_{s,\pm}(x)=\sum_{j=0}^{\pm T/2} \L_1^s(S^jx)\label{14.5}\ee
and select a point $\Bk(\Bx)\in E_{\Bx}$ for each $\Bx$. Then the SRB
distribution $\m_{SRB}$ and the volume distribution $\m_L$ on the
phase space $\O$, which we suppose to have Liouville volume, footnote
p.\pageref{2}, $V(\O)$, attribute to the {\it nonempty} sets
$E_{\V\x}$ the respective probabilities $\m$ and $\m_L$

\be \m(\Bx)\defi\m_{SRB}(E_{\Bx})\qquad \hbox{\rm and
respectively}\qquad \m_{L}(\Bx)\defi \frac{V(E_{\Bx})}{V(\O)}\label{14.6} \ee
if $V(E)$ denotes the Liouville volume of $E$. The distributions
$\m,\m_L$ are shown, \cite{GBG04,Ga00}, to be defined by

\be 
\eqalign{\m(\Bx)\,=&\,
h^T_{u,u}(\Bx)\cdot
e^{\big(-U_{u,-}^{T/2}(\k(\Bx))-U^{T/2}_{u,+}(\k(\Bx))\big)}\cr
\m_{L}(\Bx)\,=&\,
h^T_{s,u}(\Bx)\cdot 
e^{\big(U_{s,-}^{T/2}(\k(\Bx))-U_{u,+}^{T/2}(\k(\Bx))\big)}\cr}\label{14.7}
\ee
where $\k(\Bx)\in E_\Bx$ is the center of $P_{\x_0}$
and $h^T_{u,u}(\Bx)$, $h^T_{s,u}(\Bx)$ are suitable functions of
$\Bx$, {\it uniformly bounded} as $\Bx$ and $T$ vary and which are mildly
dependent on $\Bx$; so that they can be regarded as constants for the
purpose of the present discussion, {\it cfr.} Ch. 9 in
\cite{Ga00}.

If $\g$ is a scale below which all interesting observables are (for
practical purposes) constant, then choosing $T=O(\log\g^{-1})$ the
sets $E_\Bx$ are a coarse graining of phase space suitable for
computing time averages as weighted sums over the elements of the
partition. 

And both in equilibrium and out of equilibrium the SRB
distribution {\it will not attribute equal weight} to the sets
$E_\Bx$. The weight will be instead proportional to
$e^{\big(-U_{u,-}^{T/2}(\k(\Bx))-U^{T/2}_{u,+}(\k(\Bx))\big)}$, {\it
i.e.}  to the inverse of the exponential of the expansion rate of the
map $S^T$ along the unstable manifold and as a map of
$S^{-\frac{T}2}\k(\Bx)$ to $S^{\frac{T}2}\k(\Bx)$.  The more unstable
the cells are the less weight they have. Given Eq. (\ref{14.7}) the
connection with the Gibbs state with potential energy $A(\Bx)=\L_1^u(\Bx)$
appears, see \cite[Sec.4.3 and Ch. 5,6]{GBG04}.

The sets $E_{\Bx}$ represent macroscopic states, being just small
enough so that the physically interesting observables have a constant
value within them; and we would like to think that they provide us
with a model for a ``{\it coarse grained}'' description of the
microscopic states. The notion of coarse graining is, here,
precise and, nevertheless, quite flexible because it contains a free
``resolution parameter'' $\g$. Should one decide that the resolution
$\g$ is not good enough because one wants to study the system with
higher accuracy then one simply chooses a smaller $\g $ (and,
correspondingly, a larger $T$).

\def\SEC{A2: SRB and Coarse Graining: a  physicist's view}
\section{A2: SRB and Coarse Graining: a  physicist's view}
\label{secA2}\iniz

How can the analysis of Appendix A1 be reconciled with the numerical
simulations, and with the naive view of motion, as a permutation of
cells?  The phase space volume will generally contract with time: yet
we want to describe the evolution in terms of an evolution permuting
microscopic states. Also because this would allow us to count the
microscopic states relevant for a given stationary state of the system
and possibly lead to extending to stationary nonequilibria Boltzmann's
definition of entropy.

Therefore we divide phase space into {\it equal} parallelepipedal {\it
microcells} $\D$ of side size $\e\ll \g $ and try to discuss time
evolution in terms of them: we shall call such cells ``microscopic''
cells, as we do not associate them with any particular observable;
they represent the highest microscopic resolution.

The new microcells should be considered as realizations of objects
alike to those arising in computer simulations: in simulations the
cells $\D$ are the ``digitally represented'' points with coordinates
given by a set of integers and the evolution $S$ is a {\it program} or
{\it code} simulating the solution of equations of motion suitable for
the model under study. The code operates {\it exactly} on the
coordinates (the deterministic round offs, enforced by the
particular computer hardware and software, should be considered part
of the program).  

The simulation will produce (generically) a chaotic evolution ``for
all practical purposes'', {\it i.e.} 
\\ 
(1) if we only look at ``macroscopic observables'' which are constant
on the coarse graining scale $\g=e^{-\frac12\lis\l T}\ell_0$ of the
partition ${\cal P}^T$, where $\ell_0$ is the phase space size and
$\lis\l>0$ is the least contractive line element exponent (which
therefore fixes the scale of the coarse graining, by the last
definition);\footnote{\nota Here it is essential that the CH holds,
  otherwise if the system has long time tails the analysis becomes
  much more incolved and so far it can be dealt, even if only
  qualitatively, on a case by case basis.} and
\\
(2) if we look at phenomena on time scales far shorter than the
recurrence times (always finite in finite representations of motion,
but of size usually so long to make the recurrence phenomenon
irrelevant).\footnote{\nota To get an idea of the orders of magnitude
  consider a gas of $N$ particles of density $\r$ at temperature $T$:
  the metric on phase space will be $ds^2=\sum_i(\frac{d\V p_i^2}{k_B
    T}+\frac{d\V q_i^2}{\r^{-2/3}})$; hence the size of a microcell
  will be $\sqrt{O(N)}\,\d_0$ if $\d_0$ is the precision with which
  the coordinates are imagined determined (in simulations $\d_0\simeq
  10^{-14}$ in double precision) as all contributions to $ds^2$ are
  taken of order $O(1)$. Coarse grained cells contain, in all
  proposals, many particles, $O(N)$, so that their size will contain a
  factor $\d$ rather than $\d_0$ and will be $\d/\d_0=O(N^{1/3})$
  larger.} 

The latter conclusion can be reached by realizing that \*

\0(a) there has to be a small enough division into microcells that
allows us to describe evolution as a map  (otherwise
numerical simulations would not make sense);

\0(b) however the evolution map cannot be, in general, a
permutation. In simulations it will happen, {\it essentially always},
that it ({\it i.e.}  the software program) will send two distinct
microcells into the same one. It does certainly happen in
nonequilibrium systems in which phase space contracts in the
average;\footnote{\nota With extreme care it is sometimes, and in
equilibrium, possible to represent evolution with a code which is a
true permutation: the only example that I know, dealing with a
physically relevant model, is in \cite{LV93}.}

\0(c) even though the map will not be one-to-one, nevertheless it will
be such {\it eventually}: because any map on a finite space is a
permutation of the points which are recurrent. This set is the {\it
attractor} of the motions, that we call $\AA$ and which will be
imagined as a the collection of microcells approximating the unstable
manifold and intersecting it. All such microcells will be considered
taking part in the permutation: but this is not an innocent assumption
and in the end is the reason why the SRB is unique, see
below.

\0(d) every permutation can be decomposed into cycles: each cycle will
visit each coarse cell with the same frequency (unless there are more
than one stationary distributions describing the asymptotics of a set
of microcells initially distributed uniformly, a case that we exclude
because of the transitivity assumption). Hence it is not restrictive to
suppose that there is only one cycle (``ergodicity'' on the
attractor).  \*

{\it Then} consistency between the expansion of the unstable
directions and the existence of a cyclic permutation of the microcells
in the attractor $\AA$ {\it demands that the number of microcells in
each coarse grained cell $E_\Bx$, Eq.(\ref{14.3}), must be inversely
proportional to the expansion rate}, {\it i.e.} it has to be given by
the first of Eq. (\ref{14.7}).

\eqfig{200}{60}{
\ins{-20}{40}{$\EE(\Bx)$}}{cell.eps}{}

\0{\nota Fig.4: A very schematic and idealized drawing of the
  attractor layers $\st\D(\Bx)$, remaining after a transient time,
  inside a coarse cell $\st\EE(\Bx)$; the second drawing (indicated by
  the arrow) represents schematically what the layers really are, if
  looked closely: namely collections of microcells laying uniformly on
  the attractor layers, {\it i.e.} the discretized attractor
  intersected with the coarse cell.}  \*
More precisely we imagine, developing a heuristic argument, that
the attractor in each coarse cell $\EE(\Bx)$ will appear as a stack of
a few portions of unstable manifolds, the ``layers'' of 
footnote$^{\ref{footnote9}}$, whose union form the (disconnected)
surface $\D(\Bx)$ intersection between $\EE(\Bx)$ and the
attractor. Below $\D(\Bx)$ will be used to denote both the set and its
surface, as the context demands. The stack of connected surfaces
$\D(\Bx)$ is imagined covered uniformly by $N(\Bx)$ microcells, see Fig.4.

Let $t\defi T+1$. Transitivity implies that there is a smallest
integer $m\ge0$ such that $S^{t+m}\EE(\Bx)$ intersects all other
$\EE(\Bx')$: the integer $m$ is $t$-independent (and equal to the
minimum $m$ such that $Q^m_{\s,\s'}>0$). In $t+m$ time steps each
coarse cell will have visitied all the others and the layers
describing the approximate attractor in a single coarse cell will have
been expanded {\it to cover the entire attractor} for the map
$S^{t+m}$.%
\footnote{\nota To see this it is convenient to remark that the
$S^{t+m}$-image of a layer $\d(x)\subset \D(\Bx)$ of the attractor
will cover some of the layers of $\D(\Bx)$, because $S^t\EE(\Bx)$
visits and fully covers all coarse cells $\EE(\Bx')$, see
footnote$^{\ref{footnote9}}$.  Hence $S^{t+m}\D(\Bx)$ will fully cover
at least part of the layers of the attractor in $\EE(\Bx)$. Actually
{\it it will cover the whole of} $\D(\Bx)$, because if a layer of
$\D(\Bx)$ was left out then it will be left out of all the iterates of
$S^{t+m}$ and a nontrivial invariant subset of the attractor for $S^t$
would exist.} The latter coincides with the attractor for $S$ because
$S^j$ is transitive for all $j$ if it is such for $j=1$ and this
property has to be reflected by the discretized dynamics at least if
$j$ is very small compared to the (enormous) recurrence time on the
discrete attractor as $t$ is, being a time on the coarse grain scale.

Suppose first that $m=0$, hence $S^t\D(\Bx)$ is the entire attractor
for all $\Bx$. This is an assumption useful to exhibit the idea but
unrealistic for invertible maps: basically this is realized in the
closely related SRB theory for a class of non invertible expansive
maps of the unit interval).

So the density of microcells will be $\r(\Bx)=\frac{N(\Bx)}{\D(\Bx)}$
and under time evolution $S^t$ the unstable layers $\D(\Bx')$ in
$\EE(\Bx)$ expand and cover all the layers in the cells $\EE(\Bx')$.
If the coarse cell $\EE(\Bx)$ is visited, in $t=T+1$ time steps, by
points in the coarse cells $\Bx'$, a property that will be
symbolically denoted $\EE_{\Bx'}\in S^{-t}\EE(\Bx)$, a fraction
$\n_{\Bx,\Bx'}$ of the $N(\Bx')$ microcells will end in the coarse
cell $\EE(\Bx)$, and $\sum_{\Bx}\n_{\Bx,\Bx'}=1$. Then consistency
with evolution as a cyclic permutation demands 

\be N(\Bx)=\sum_{\Bx'}
\frac{N(\Bx')}{\D(\Bx')}\frac1{e^{\L_{u,T}(\Bx')}}\D(\Bx)
\defi\LL(N)(\Bx),\qquad\hbox{\it i.e.}
\label{15.1}\ee
because the density of the microcells on the images of $\D(\Bx')$
decreases by the expansion factor $e^{\L_{u,T}(\Bx')}$, so that
$\n_{\Bx,\Bx'}= \frac{\D(\Bx)}{\D(\Bx')}\frac1{e^{\L_{u,T}(\Bx')}}$.

As a side remark it is interesting to point out that for the density
$\r(\Bx)$ Eq.(\ref{15.1}) becomes simply $\r(\Bx)=\sum_{\Bx'}
e^{-\L_{u,T}(\Bx')}\r(\Bx')$, closely related to the similar equation
for invariant densities of Markovian surjectiive maps of the unit
interval, \cite{GBG04}.

The matrix $\LL$ has all elements $>0$ (because $m=0$) and therefore
has a simple eigenvector $v$ with positive components to which
corresponds the eigenvalue $\l$ with maximum modulus: $v=\l\,\LL(v)$
(the ``Perron-Frobenius theorem'') with $\l=1$ (because
$\sum_{\Bx}\n_{\Bx,\Bx'}=1$). It follows that the consistency
requirement uniquely determines $N(\Bx)$ as proportional to $v_\Bx$.
Furthermore $S^{t}\D(\Bx)$ is the entire attractor; then its surface
is $\Bx$ independent and equal to $e^{\L_{u,T}(\Bx)}\D(\Bx)$:
therefore $N(\Bx)=\,const\, e^{-\L_{u,T}(\Bx)}$.
 
The general case is discussed by considering $S^{t+m}$ instead of
$S^{t}$: this requires taking advantage of the properties of the
ratios $e^{\L_{u,T}(\Bx)}/e^{\L_{u,T+m}(\Bx)}$. Which are not only 
uniformly bounded in $T$ but also only dependent on the sequence
$\Bx=(\x_{-\frac12T},\ldots,\x_{\frac12T})$ through a few symbols with labels 
near $-\frac12T$ and $\frac12T$: this correction can be considered part of the 
factors $h^T_{u,u}$ in the rigorous formula Eq.(\ref{14.7}).

Note that $e^{\L_{u,T}(\Bx)}\D(\Bx)=$ constant reflects Pesin's
formula, \cite{GBG04}, for the approximate dynamics considered here.
\*

So the SRB distribution arises naturally from assuming that dynamics
can be discretized on a regular array of point (``microcells'') and
become a one cycle permutation of the microcells on the
attractor. This is so under the CH and holds whether the dynamics is
conservative (Hamiltonian) or dissipative.

\*

\0{\it Remark:} It is well known that hyperbolic systems admit
(uncountably) many invariant probability distributions, besides the
SRB. This can be seen by noting that the space of the configurations
is identified with a space of compatible sequences. On such a space
one can define uncountably many stochastic processes, for instance by
assigning an arbitrary short range translation invariant potential,
and regarding the corresponding Gibbs state as a probability
distribution on phase space.  However the analysis just presented
apparently singles out SRB as the unique invariant distribution. This
is due to our assumption that, in the discretization, microcells are
regularly spaced and centered on a regular discrete lattice and
evolution eventually permutes them in a (single, by transitivity)
cycle consisting of the microcells located on the attractor (and
therefore locally evenly spaced, as inherited from the regularity of
the phase space discretization).  

\0Other invariant distributions can be obtained by custom made
discretizations of phase space which will not cover the attractor in a
regular way.  This is what is done when other distributions, ``not
absolutely continuous with respect to the phase space volume'', are to
be studied in simulations. A paradigmatic example is given by the map
$x\to 3x \, {\rm mod}\, 1$: it has an invariant distribution
attributing zero probability to the points $x$ that, in base $3$, lack
the digit $2$: to find it one has to write a program in which data
have this property and make sure that the round off errors will not
destroy it. Almost any ``naive'' code that simulates this dynamics using double
precision reals represented in base $2$ will generate, instead, the
corresponding SRB distribution which is simply the Lebesgue measure on
the unit interval (which is the Bernoulli process on the symbolic
dynamics giving equal probability $\frac13$ to each digit).

  \*

The physical representation of the SRB distribution just obbtained, see
\cite{Ga95a,Ga00}, shows that there is no conceptual difference
between stationary states in equilibrium and out of equilibrium. In
both cases, if motions are chaotic they are permutations of
microcells and the {\it SRB distribution is simply equidistribution
over the recurrent microcells}. In equilibrium this gives the Gibbs
microcanonical distribution and out of equilibrium it gives the SRB
distribution (of which the Gibbs one is a very special case).

The above heuristic argument is an interpretation of the mathematical
proofs behind the SRB distribution which can be found in
\cite{Bo75,GBG04}, (and heuristically is a proof in itself). Once
Eq. (\ref{14.7}) is given, the expectation values of the observables
in the SRB distributions can be formally written as sums over suitably
small coarse cells and symmetry properties inherited from symmetries
of the dynamic become transparent. The Fluctuation Theorem is a simple
consequence of Eq. (\ref{14.7}), see Appendix A3: however it is
conceptually interesting because of the surprising unification of
equilibrium and nonequilibrium behind it.

The discrete repesentation, in terms of coarse grain cells and
microcells leads to the possibility of counting the number $\NN$ of
the microcells and therefore to define a kind of entropy function: see
\cite{Ga01} where the detailed analysis of the counting is performed
and the difficulties arising in defining an entropy function as
$k_B\log \NN$ are critically examined.

\section{A3: Why does FT hold?}
\def\SEC{A3: Why FT?}\label{secA3}\iniz

As mentioned the proof of FT in quite simple, \cite{GC95}. By the
first of Eq. (\ref{14.5}), (\ref{14.7}) and by the theory of
$1D$-short range Ising models, see \cite{Ga95b} for details, the
probability that $p$ is in a small interval centered at $p$ compared
to the probability that it is in the opposite interval is

\be
\frac{P_\t(p)}{P_\t(-p)}=\frac{\sum_{i \to
    p\s_+\t}e^{-\sum_{-\t/2}^{\t/2} \L^u_{1}(S^k \k_i)+B(i,\t)}}
{\sum_{i \to
    -p\s_+\t}e^{-\sum_{-\t/2}^{\t/2} \L^u_{1}(S^k \k_i) +B(i,\t)}}
\label{16.1}\ee
where $\sum_{i \to p\s_+\t}$ is sum over the centers $\k_i$ of the
rectangles $E_i$ labeled by $i\defi(\x_{-\t/2},\ldots,\x_{\t/2})$ with
the property

\be \sum_{k=-\t/2}^{\t/2}
\s(S^k\k_i)+B(i,\t)\simeq p\s_+\t\,\label{16.2}\ee
where $\simeq$ means that the left hand side is contained in a very
small interval (of size of order $O(1)$, \cite{Ga95b}, call it $b$)
centered at $p\s_+\t$; the $B(i,\t)$ is a term of order $1$ (a
boundary term in the language of the Ising model interpretation of the
SRB distribution): $|B(i,\t)|\le b<+\infty$: and it takes also into
account the adjustments to be made because of the
arbitrariness of the choice of $\k_i$.%
\footnote{Which is taken here $\Bk_i=\,$the center of $P_{\x_0}$, but
  which could equivalently made by choosing other points in $E_\Bx$,
  for instance by continuing the string
  $i=(\x_{-\t/2},\ldots,\x_{\t/2})$ to the right and to the left,
  according to an a priori fixed rule depending only on $\x_{\t/2}$
  and $\x_{-\t/2}$ respectively. Thus turning it to a biinfinite
  compatible string $\Bx_i$ which therefore fixes a new point
  $\Bk'_i$.} Independence on $i,\t$ of the bound on $B(i,\t)$
  reflects smoothness of $S$ and elementary properties of
  short range $1D$ Ising chains, \cite{Ga95b}.

Suppose that the symbolic dynamics has been chosen time reversible,
{\it i.e.} the time reversal map $I$ maps $P_i$ into $IP_i=P_{I(i)}$
for some $I(i)$: this is not a restriction as discussed in Appendix A1.
Then the above ratio of sums can be rewritten as a ratio of sums over
the same set of labels,

\be
\frac{P_\t(p)}{P_\t(-p)}=\frac{\sum_{i \to
    p\s_+\t}e^{-\sum_{-\t/2}^{\t/2} \L^u_{1}(S^k \k_i)+B(i,\t)}}
{\sum_{i \to
    p\s_+\t}e^{-\sum_{-\t/2}^{\t/2} \L^u_{1}(S^k I(\k_i)) +B(I(i),\t)}}.
\label{16.3}\ee
Remark that $\L^u_1(Ix)=-\L^s_1(x)$ (by time reversal symmetry) and
that (by Eq. (\ref{14.3})) $ \sum_{k=-\t/2}^{\t/2}
(\L^u_1(S^k(x))+\L^s_1(S^{-k}(x)))$ can be written as

\be
\sum_{k=-\t/2}^{\t/2}
(\L^u_1(S^k(x))+\L^s_1(S^{k}(x)))=\sum_{k=-\t/2}^{\t/2}
\s(S^kx)+B(x,\t)\label{16.4}\ee
with $B(x,\t)\le b$ (again by the smoothness of $S$), possibly
redefining $b$.

Therefore the ratio of corresponding terms in the numerator and
denominator ({\it i.e.} terms bearing the same summation label $i$) is
precisely $p\s_+\t$ up to $\pm3b$. Hence

\be  e^{\t\s_+\,p\,-3b}\le \frac{P_\t(p)}{P_\t(-p)}< e^{\t\s_+\,p\,+3b}
\label{16.5}\ee
so that FT holds for finite $\t$ with an error $\pm\frac{3b}\t$,
infinitesimal as $\t\to+\infty$. For a detailed discussion of the
error bounds see \cite{Ga95b}. 

{\it Of course} for all this to make sense the value of $p$ must be
among those which not only are possible but also such that the values
close enough to possible values are possible. This means that $p$ has
to be an internal point to an interval of values that contains limit
points of $\lim_{\t\to+\infty} \frac1\t\sum_{k=0}^{\t} \frac{\s(S^k
x)}{\s_+}$ for a set of $x$'s with positive SRB probability: the value
$p^*$ in FT is the supremum among the value of $p$ with this property,
\cite{Ga95b} (contrary to statements in the literature this physically
obvious remark is explicitly present in the original papers: and one
should not consider the three contemporary references,
\cite{GC95,Ga95b,GC95b}, has having been influenced by the doubts on
this point raised much later.)

The assumptions have been: (a) existence of a Markovian partition,
{\it i.e.} the possibility of a well controlled symbolic dynamics
representation of the motion; (b) smooth evolution $S$ and (c) smooth
time reversal symmetry: the properties (a),(b) are equivalent to the
CH. Of course positivity of $\s_+$ is {\it essential}, in spite of
contrary statements; if $\s_+=0$ the leading terms would come
from what has been bounded in the remainder terms and, in any event
the analysis world be trivial, with or without chaoticity assumptions,
\cite{BGGZ05}.

Since Lorenz, \cite{Lo63}, symbolic dynamics is employed to represent
chaos and many simulations make currently use of it; smoothness has
always been supposed in studying natural phenomena (lack of it being
interpreted as a sign of breakdown of the theory and of necessity of a
more accurate one); time reversal is a fundamental symmetry of nature
(realized as $T$ or $TCP$ in the Physics notations). Hence in spite of
the ease in exhibiting examples of systems which are not smooth, not
hyperbolic, not time reversal symmetric (or any subset thereof) the CH
still seems a good guide to understand chaos.

\section{A4: Harmonic Thermostats}
\def\SEC{A4: Harmonic Thermostats}\label{secA4}\iniz

Here the ``efficiency'' of a harmonic thermostat is discussed. It
turns out that in general a thermostat consisting of infinite free
systems is a very simple kind of Hamiltonian thermostat, but it has to
be considered with caution as it can be inefficient in the sense that
it might not drive a system towards equilibrium (i.e. towards a Gibbs
distribution). In the example given below a system in interaction with
an infinite harmonic reservoir at inverse temperature $\b$ is
considered.
It is shown that the interaction can lead to a stationary state, of
the system plus reservoir, which is not the Gibbs state at temperature
$\b^{-1}$. The following is a repetition of the analysis in
\cite{ABGM72}, adapting it to the situation considered here.

A simple model is a $1$-dimensional harmonic oscillators chain, of
bosons or fermions, initially in a Gibbs state at temperature
$\b^{-1}$. The Hamiltonian for the equilibrium {\it initial} state
will be

\be  H_0=\sum_{x=1}^{N-1}
-\frac{\hbar^2}{2m}\D_{q_x}+\sum_{x=1}^{N-1}\frac{m\o^2}2
q_x^2+\sum_{x=1}^N\frac{m\m^2}2 (q_x-q_{x-1})^2\label{17.1}\ee
with boundary conditions $q_0=q_N=0$ and $\hbar,m,\o^2,\m^2>0$.  The
initial state will be supposed to have a density matrix
$\r_0=\frac{e^{-\b H_0}}{{\rm Tr}\, e^{-\b H_0}}$. Time evolution will
be governed by a {\it different} Hamiltonian

\be  H_\l=H_0+\frac{m\l}2 q_1^2,\qquad \l+\o^2>0\label{17.2}\ee
The question of ``thermostat efficiency'' is: does $\r_t\defi
    e^{\frac{i}{\hbar} t H_\l}\r_0 e^{-\frac{i}{\hbar} t H_\l}$ converge
    as $t\to+\infty$ to $\r_\infty=\frac{e^{-\b H_\l}}{{\rm Tr}\, e^{-\b
    H_\l}}$. Or: does the system consisting in the oscillators labeled
    $2,3,\ldots$ succeed in bringing up to the new equilibrium state
    the oscillator labeled $1$?
Convergence means that the limit $\media{A}_{\r_t}\tende{t\to+\infty}$
$\media{A}_{\r_\infty}$ exists, at least for the observables $A$
essentially localized in a finite region.

The Hamiltonian in Eq.(\ref{17.2}) can be diagonalized by studying the 
matrix

\be 
V_\l=m\,\pmatrix{\o^2+2 \m^2+\l& -\m^2&0         &\ldots\cr
                -\m^2    &\o^2+ 2\m^2 &-\m^2 &\ldots\cr
                      0      &-\m^2 &\o^2+2 \m^2&\ldots\cr
                  \ldots&\ldots&\ldots&\ldots}\defi V_0+\l m P_1\label{17.3}\ee
The normalized eigenstates and respective eigenvalues of $V_0$ are

\be 
\Ps^0_k(x)\defi\sqrt{\frac{2}{N}}\,\sin\frac{\p k}Nx,\qquad
  \L^0_{k}=m\,\Big(\o^2+{2\m^2(1-\cos\frac{\p k}N)}\Big)\label{17.4}\ee
and the vectors $\Ps^0_k$ will be also denoted $\ket{k}$ or $\ket{\Ps^0_k}$.  

To solve the characteristic equation for $V_\l$, call $\Ps$ a generic
normalized eigenvector with eigenvalue $\L$; the eigenvalue equation is

\be \bra{k}\ket\Ps (\L^0_{k}-\L)+\l\,m\,\bra k
\ket\BO\,\bra\BO\ket\Ps=0\label{17.5}\ee
where $\BO$ is the vector $\BO=(1,0,\ldots,0)\in\CC^{N-1}$, so that
$P_1=\ket\BO \brav\BO$. Hence, noting that $\bra{\BO}\ket{\Ps}$ cannot
be $0$ because this would imply that $\L=\L^0_{k}$ for some $k$ and
therefore $\ket{\Psi}=\ket{k}$ which contradicts $\bra\BO\ket\Ps=0$,
it is

\be \bra{k}\ket\Ps=-\l\,m\,
\frac{\bra{k}\ket\BO\cdot\bra\BO\ket\Ps}{\L^0_{k}-\L}\label{17.6}\ee
and the compatibitity condition that has to be satisfied is

\be 
\frac{\bra\BO\ket\Ps}{\l\,m}\,=
\,\sum_{k=1}^{N-1}\frac{|\bra\BO\ket{k}|^2}{\L-\L^0_{k}}
\bra\BO\ket\Ps=
\,\sum_{k=1}^{N-1}\frac{2\sin^2\frac{\p
    k}{N}}{N}\frac{\bra\BO\ket\Ps}{\L-\L^0_{k}}.\label{17.7}\ee
Once Eq.(\ref{17.7}) is satisfied,  Eq.(\ref{17.6}) imply that
the eigenvalue equation, Eq.(\ref{17.5}), is satisfied, and by a
$\ket\Psi\ne0$ (determined up to a factor).

The Eq.(\ref{17.7}) has $N-1$ solutions, corresponding to the $N-1$
eigenvalues of $V_\l$. This follows by comparing the graph of
$y(\L)\equiv\frac1{\l m}$ with the graph of the function of $\L$ in the
intermediate term of Eq.(\ref{17.7}).  One of the solutions remains
isolated in the limit $N\to\infty$, because the equation

\be 1=\frac{2\,\l\,m}{\p}\int_0^\p\frac{\sin^2\k}{\L-\L^0(\k)}
d\k,\qquad \L^0(\k)\defi
m\,\Big(\o^2+4\m^2\sin^2\frac{\k}2\Big)\label{17.8}\ee
has, uniformly in $N$, only one isolated solution for $\L<\inf \L^0(\k)=m
\o^2$ if $\l<0$, or for $\L>\sup \L^0(\k)$ if $\l<0$. Suppose for
definiteness that $\l<0$.

Let $\Ps^\l_k(x),\,k=1,\ldots,N-1$, be the corresponding
eigenfunctions. The matrices $ U_{\l;k,x}=\Ps^\l_k(x)$ are unitary and
$(U_\l)_{\l=0}\equiv U_0$. It is $U_{0;k,x}=\sqrt{\frac2{N}}\sin
\frac{\p k}N x$ and 
$\bra{\Ps^0_k}\ket{\Ps^\l_{k'}}=\frac{\bra{k}\ket\BO}{Z_N(k')(\L^\l_{k'}-\L^0_k)}$
with $Z_N(k')^2=\sum_{k}
\frac{|\bra{k}\ket\BO|^2}{(\L^\l_{k'}-\L^0_k)^2}$ by Eq.(\ref{17.6}).
Then setting $\a^\pm_x= \frac{p_x\pm
iq_x}{\sqrt2}$ let

\be
 a^+_{\l;k} \defi (U_\l \Ba^+)_k, \qquad a^-_{\l;k}=\defi
 (\Ba^-U_\l^*)_k
\label{17.9}\ee
where $U^*$ is the adjoint of $U$ (so that $UU^*=1$ if $U$ is unitary).
It is

\be \a^+_{x}=\sum_k \lis U_{\l;k,x} a^+_{\l;k},\qquad a^+_{\l;k }=
\sum_{h,y} U_{\l;k,y}\lis U_{0;h,y} a^+_{0;h}
\label{17.10}\ee
if the overbars denote complex conjugation.

The operators $a^\pm_{\l,k}$ will be creation and
annihilation operators for quanta with energy $\hbar
\sqrt\frac{\L^\l_k}{m}\defi E_\l(k)$. So a state with $n_k=0,1,\ldots$
quanta in state $k$ will have energy $\sum_k E_\l(k)(n_k+\frac12)$.

Consider the observable $a^+_{\l,1}a^-_{\l,1}=A$. Its average is
{\it time independent}, in the evolution generated by $H_\l$, and if
$W\defi U_\l U^*_0$ it is equal to

\be
\eqalign{&\media{A}_{\r_t}\equiv \media{A}_{\r_0}\equiv {\,\rm Tr\,} 
\r_0 (W \V a^+_0)_1 (W \V a^-_0)_1\cr
&=\sum_k {\rm Tr\,}\r_0W_{1,k} W_{1,k'} a^+_{0,k}a^-_{0,k'}
=\sum_{k=1}^{N-1}|W_{1,k}|^2 \frac{\sum_{n=0}^{n_f} e^{-\b E_0(k)n}\,n}
{\sum_{n=0}^{n_f} e^{-\b E_0(k)n}}
\cr}\label{17.11}\ee
where $n_f=1$ if the statistics of the quanta is fermionic (this was
the case in \cite{ABGM72}) or $n_f=+\infty$ if it is bosonic. In the
two cases the result is

\be \sum_k|W_{1,k}|^2\frac1{e^{\b E_0(k)}\pm1}\label{17.12}\ee
If the system reached thermal equilibrium, setting
$\r_\l(k)\defi\frac1{e^{\b E_\l(k)}\pm1}$, this should be $\r_\l(1)$,
which is impossible, as it can be checked by letting $\b\to+\infty$
and remarking that it is $E_\l(1)<E_0(1)$ with a difference positive
uniformly in $N$.
Furthermore the observable $A$ is localized near the site $x=1$:
because the wave function of the lowest eigenvalue is
$\frac1{Z_N(1)}\sum_h
\frac{\bra{h}\ket\BO}{\L^0_h-\L^\l_k}\ket{\Ps^0_h}$ so that

\be\Ps^\l_1(x)=\frac1{Z_N(1)}\sum_h\frac{\Ps^0_h(1)\Ps^0_h(x)}{\L^\l_1-\L^0_h}
\tende{N\to\infty}\frac 1{Z_\infty}
\frac2\p\int_0^\p\frac{\sin\k\,\sin\k
  x}{\L^\l_1-\L^0(\k)}\,d\k\label{17.13}
\ee
and the integral tends to $0$ as $x\to\infty$ faster than any
power, so that $0<Z_\infty<\infty$
and $\Ps^\l_1$ is normalizable.

Therefore the thermostatic action of the system in
the sites $2,3,\ldots$ on the site $1$ is {\it not efficient} and the
state does not evolve towards the Gibbs state at temperature
$\b^{-1}$, not even in the limit $N\to+\infty$.

This result should be contrasted with the closely related case in
which the system oscillator at $1$ plus the others is started in a
equilibrium state for $H_\l$ and at time $0$ is evolved with
Hamiltonian $H_0$. In this case the system thermalizes properly, see
the analogous analysis in \cite{ABGM72}, see also \cite{HL73b} for a
large class of related examples.

Of course the question of effectiveness of a thermostat could be
discussed also for non linear theormostats, finite or infinite. It
seems that, under mild assumptions, non linear thermostat models should
be efficient, {\it i.e.} generate proper heat exchanges even when
acting only at the boundary as in the case of the thermostats
considered in Sec.\ref{sec9}. The analysis in \cite{GG07} gives some
preliminary evidence in this direction.

Harmonic thermostats are nevertheless very interesting, provided the
above pathologies are excluded by a careful formulation of the models:
see for instance \cite{HL73b}, see also \cite{HI05}. It is also clear
that the pathologies seem to be related to the fact that the
thermostats constituents are ``not interacting'' or ``linearly
interacting'': their origin in the above analysis is shown to be
related to the existence of isolated eigenvalues of the Hamiltonian at
the bottom of the spectrum and this is the property that should be
excluded.  The pathologies are likely to be absent in models in which
there is nonlinear interaction within the thermostats constituents so
that such models should be perfectly well behavng ({\it i.e.}
efficient in the sense of this paper).  However the latter models are
also highly nontrivial even at a purely mathematical level.

\def\SEC{A5: Bohmian Quantum Systems}
\section{A5: Bohmian Quantum Systems}
\label{secA5}\iniz

Consider the system in Fig.1 and suppose, as in Sec.\ref{sec10},
that the nonconservative force $\V E(\V X_0)$ acting on the system
vanishes, {\it i.e.} consider the problem of heat flow through
$\CC_0$.  Let $H$ be the operator on $L_2(\CC_0^{3N_0})$, space of
symmetric or antisymmetric wave functions $\Ps$,

\be H_{\V X}=
-\frac{\hbar^2}{2m}\D_{\V X_0}+ U_0(\V X_0)+\sum_{j>0}\big(U_{0j}(\V X_0,\V
X_j)+U_j(\V X_j)+K_j\big)\label{18.1}\ee
where $\D_{\V X_0}$ is the Laplacian, and note that its spectrum
consists of eigenvalues $E_n=E_n(\{\V X_j\}_{j>0})$, depending on the
configuration $\V X\defi\{\V X_j\}_{j>0}$,

Thermostats will be modeled as assemblies of classical particles as
in Sec.\ref{sec9}: thus their temperature can be defined as the
average kinetic energy of their particles and the question
of how to define it does not arise.

The viewpoint of Bohm on quantum theory seems quite well adapted to
the kind of systems considered here.  A system--reservoirs model
can be the {\it dynamical system} on the variables
$\big(\Ps,\V X_0,(\{\V X_j\},$ $\{\V{{\dot X}}_j\})_{j>0}\big)$
defined by

\be \eqalign{
-i\hbar {\dot\Ps(\V X_0)}=& \,( H_{\V X}\Ps)(\V X_0),\cr
\V{{\dot X}}_0=&\,\hbar\,%
\Im\,\,\frac{\BDpr_{\V X_0} \Ps(\V X_0)}{\Ps(\V X_0)},
\kern4mm{\rm and\ for}\
j>0\cr
\V{{\ddot X}}_j=&-\Big(\partial_j U_j(\V X_j)+
\partial_j U_j(\V X_0,\V X_j)\Big)-\a_j \V{{\dot X}}_j\cr
\a_j\defi&\frac{W_j-\dot U_j}{2 K_j}, \qquad
W_j\defi -\V{{\dot X}}_j\cdot \V\partial_j U_{0j}(\V X_0,\V
X_j)\cr
}\label{18.2} \ee
here the first equation is Schr\"odinger's equation, the second is the
vlocity of the Bohmian particles carried by the wave $\Ps$, the others are
equations of motion for the thermostats particles analogous to the one
in Eq.(\ref{9.1}), (whose notation for the particles labels is
adopted here too). Evolution maintains the thermostats kinetic
energies $K_j\equiv \frac12\V{{\dot X}}_j^2$ exactly constant so that
they will be used to define the thermostats temperatures $T_j$ via
$K_j=\frac32 k_B T_j N_j$, as in the classical case.

Note that if there is no coupling between system and thermostats, {\it
i.e.} the system is ``isolated'', then there are many invariant
distributions: {\it e.g.} the probability distributions $\m$
proportional to

\be 
\sum_{n=1}^\infty e^{-\b_0
  E_n}\d(\Ps-\Ps_n\,e^{i\f_n})\,\big|\Ps(\V X_0)|^2
 {d\f_n}d\V X_0\,\prod_j\d(\dot{\V X}_j^2-2K_j) d\dot{{\V
     X_j}}d\V X_j
\label{18.3}\ee
where $E_n$ and $\Ps_n$ are time independent, under the assumed
absence of interaction between system and thermostats, and are the
eigenvalues and the correspoding eigenvectors of $H$.  Then the
distributions $\m$ are invariant under the time evolution.

Time invariance of this kind of distributions is discussed in
\cite[Sec.4]{DGZ92}, where it appears as an instance of what is called
there a ``quantum equilibrium''.  The average value of an observable
$O(\V X_0)$, which depends only on position $\V X_0$, will be the
``usual'' Gibbs average

\be \media{O}_\m=Z^{-1}\int {\rm Tr}\,(e^{-\b_0 H}O))\label{18.4}\ee

For studying nonequilibrium stationary states consider several
 thermo\-stats with interaction energy with $\CC_0$, $W_j(\V X_0,\V
 X_j)$, as in Eq. (\ref{9.1}).  The equations of motion should be
 Eq. (\ref{18.2})

In general solutions of Eq.(\ref{18.2}) {\it will not be quasi periodic} and
the Chaotic Hypothesis, \cite{GC95b,Ga00,Ga07}, can be assumed: if so the
dynamics should select an invariant distribution $\m$. The
distribution $\m$ will give the statistical properties of the
stationary states reached starting the motion in a thermostat
configuration $(\V X_j,\V{{\dot X}}_j)_{j>0}$, randomly chosen with
``uniform distribution'' $\n$ on the spheres $m\V{{\dot X}}_j^2=3N_jk_B
T_j$ and in a random eigenstate of $H$. The distribution $\m$, if
existing and unique, could be named the {\it SRB distribution}
corresponding to the chaotic motions of Eq.(\ref{18.2}).

In the case of a system {\it interacting with a single thermostat} the
latter distribution should be equivalent to the canonical
distribution.  As in Sec.\ref{sec11} an important consistency check for
the model just proposed in Eq.(\ref{18.2}) is that there should exist
at least one stationary distribution $\m$ equivalent to the canonical
distribution at the appropriate temperature $T_1$ associated with the
(constant) kinetic energy of the thermostat: $K_1=\frac32 k_B
T_1\,N_1$. However also in this case, as already in Sec.\ref{sec11}, it
does not seem possible to define a simple invariant distribution, not
even in the adiabatic approximation. As in Sec.\ref{sec11}, equivalence
between $\m$ and a Gibbs distribution at temperature $T_1$ can only be
conjectured.

Furthermore it is not clear how to define phase space contraction,
hence how to formulate a FT, although the equations are reversible.

\def\SEC{References}\iniz
\label{secRef}
{\nota\baselineskip=9pt
\ifnum\biblio=0\bibliography{0Bibcaos}\fi
\ifnum\biblio=1
 \fi
\bibliographystyle{unsrt}
}
%
\end{document}